\pdfoutput=1

\documentclass[aip,jcp,superscriptaddress,reprint,showkeys]{revtex4-1}

\usepackage[english]{babel}
\usepackage[utf8]{inputenc}
\usepackage[T1]{fontenc}

\usepackage{graphicx}
\usepackage{color}

\usepackage{calligra}
\usepackage{amsmath}
\usepackage{mathrsfs}

\newcommand{\ac}[1]{\mbox{#1}}
\newcommand{\acs}[1]{\mbox{#1}}

\begin{document}

\author{Thomas Weike}
\email{t.weike@uni-bielefeld.de}
\author{Uwe Manthe}
\email{uwe.manthe@uni-bielefeld.de}
\affiliation{ Theoretische Chemie, Fakult\"at f\"ur Chemie, Universit\"at Bielefeld, Universit\"atsstr. 25, D-33615 Bielefeld, Germany}
\title{The multi-configurational time-dependent Hartree approach in optimized second quantization:
imaginary time propagation and particle number conservation}
\date{\today}

\begin{abstract}
The multi-layer multi-configurational time-dependent Hartree (MCTDH) in optimized second 
quantization representation (oSQR) approach combines the tensor contraction scheme of the 
multi-layer MCTDH approach with the use of an optimized time-dependent orbital basis. 
Extending the original work on the subject [Manthe, Weike, J. Chem. Phys. 146, 064117 (2017)], 
here MCTDH-oSQR propagation in imaginary time and properties related to 
particle number conservation are studied. Difference between the orbital equation of motion
in real and imaginary time are highlighted and a new gauge operator which facilitates efficient
imaginary time propagation is introduced.
Studying Bose-Hubbard models, particle number conservation in MCTDH-oSQR calculations is
investigated in detail. Interesting properties of the single-particle functions used in
the multi-layer MCTDH representation are identified. Based on these results, a tensor 
contraction scheme which explicitly utilizes particle number conservation is suggested.
\end{abstract}

\maketitle

\section{Introduction}

The (multi-layer) multi-configurational time-dependent Hartree (MCTDH)
approach\cite{MMC,MMC1,WT3,M6} facilitates accurate high-dimensional
quantum dynamics calculations.  It is used by many research groups to
study a variety of
systems.\cite{MCTDHex1,WBRSM,WestPNM,MCTDHex2,W1,MCTDHex4,MCTDHex5,
  MCTDHex6,MCTDHex7,MCTDHex8,MCTDHex9,MCTDHex10,MCTDHex11, MCTDHaB10,
  MCTDHex13,MCTDHex14,MCTDHex15,MCTDHex16,MCTDHex17,MCTDHex18}
Benchmark applications studied bimolecular
reactions\cite{HM1,HM2,WWM,SM,vHNM,NvHM,SM02,SM04,WeM5,WeM6,WeM8,WeM7,ElM2,ElM3},
proton transfer\cite{CVM,HCVM,HaM1,MAMCTDH,HaM2,MAMCTDH2}, and large
amplitude motion\cite{H5O2+MCTDH,H5O2+MCTDH2,H5O2+MCTDH3,H5O2+MCTDH4,
  H5O2+MCTDH5} for molecular systems including six to nine atoms on
accurate ab initio potential energy surfaces. \ac{MCTDH} simulations
based on model Hamiltonians describing non-adiabatic
dynamics\cite{WMC,WMC2,RWMC} or electron and proton transfer processes
in condensed matter\cite{WST,WT5,KCBWT,CTW2,Wreview} considered
significantly larger numbers of degrees of freedom ranging up to
$10^3$-$10^4$.

The \ac{MCTDH} approach was originally introduced to describe the
quantum dynamics of molecular systems. The molecular wave function was
specified as a function depending on geometry-dependent coordinates
and indistinguishable particle symmetries were not explicitly
considered. Subsequently MCTDH-type approaches designed for the
efficient description of bosonic and fermionic systems have been
introduced\cite{MCTDHF_1,MCTDHF_2,MCTDHF_3,MCTDHB_1,MCTDHB_2,MLMCTDHB,
  MCTDHSQR} and applied in different areas\cite{MCTDHBex2, MCTDHBex3,
  MCTDHBex12, MCTDHBex5, MCTDHBex6, MCTDHBex7, MCTDHBex8, MCTDHBex9,
  MCTDHBex10, MCTDHBex11, MCTDHBex12, MCTDHBex13, MCTDHBex14,
  MCTDHBex15, MCTDHBex16, MCTDHBex17, MCTDHBex18, MCTDHBex20,
  MCTDHBex21, MCTDHBex22, MCTDHBex23, MCTDHBPruned, MCTDHFex2,
  MCTDHFex3, MCTDHFex9, MCTDHFex8, MCTDHFex1, MCTDHFex4, MCTDHFex12,
  MCTDHFex10, MCTDHFex11, MCTDHFex5, MCTDHFex6, MCTDHFex7,
  MCTDHFnonBO, MCTDHBF_1, MCTDHBF_2, MCTDHBF_3, MLMCTDHX, WPHT,
  MCTDHSQRex2, MCTDHSQRex3, MCTDHSQRex4, MCTDHSQRex5,
  MCTDHSQRex6}. These approaches are often formulated in second
quantization. Two different strategies have mainly be utilized in this
context.

One family of approaches \cite{MCTDHF_1, MCTDHF_2, MCTDHF_3, MCTDHB_1,
  MCTDHB_2, MLMCTDHB, MLMCTDHX}, collectively denoted as MCTDH-X
approaches in the following, employ time-dependent orbitals and a
particle-number based truncation of the (time-dependent) configuration
space. The active space is typically restricted to include all
configurations with a specific number of particles. Consequently, the
active space employed in MCTDH-X approaches is invariant under
time-dependent unitary transformation of the orbitals. The numerical
efficiency of the approach relies on the combination of the
particle-number based truncation scheme and the use of optimized
time-dependent orbitals.

The multi-layer MCTDH approach in second quantization representation
(\ac{ML-MCTDH-SQR}) \cite{MCTDHSQR} employs a different strategy. Here
the numerical efficiency is achieved by utilizing tensor contraction
instead of truncation. Wang and Thoss introduced the approach using a
primitive time-independent orbital basis.  The ML-MCTDH-SQR approach
is designed to treat general Hamiltonians and does not consider
particle number-based properties explicitly.

Recently the multi-layer MCTDH in optimized second quantization
representation \ac{ML-MCTDH-oSQR} which includes features of the
\ac{ML-MCTDH-SQR} approach and the \ac{MCTDH-X} approaches has been
introduced.\cite{MW} The \ac{ML-MCTDH-oSQR} approach combines the use
of optimized time-dependent orbitals with a ML-MCTDH-type tensor
contraction scheme. In the ML-MCTDH-oSQR scheme the invariance with
respect to time-dependent unitary transformations of the orbital
basis, which is a typical feature in other MCTDH-type approaches, is
absent. These unitary transformations give rise to gauge operators
appearing in the MCTDH equations of motion. While the choice of these
gauge operators does not affect the convergence properties in other
MCTDH-type approaches, it is an important issue in the ML-MCTDH-oSQR
approach. The choice of a natural orbital basis was suggested in
Ref. \citenum{MW} and the corresponding gauge operators were derived. A
numerical test for the spin-boson-model was presented to illustrate
the efficiency of the approach.

The present work explores aspects of the ML-MCTDH-oSQR approach which
have not been addressed in Ref. \citenum{MW}. In this original work only
propagation in real time was considered. However, the presence of
relevant gauge operators results in significant differences between
real and imaginary time propagation in the ML-MCTDH-oSQR
approach. While gauge operators are hermitian in the real time
equations of motion, they are anti-hermitian in the imaginary time
ones. This difference affects the choice of efficient gauge
operators. A new type of gauge operators more suitable for imaginary
time propagations is introduced and tested in the present work.

Particle number conservation is another important aspect not
considered in previous work on the ML-MCTDH-SQR and ML-MCTDH-oSQR
approaches. These approaches are formulated in Fock space and thus can
contain states with all possible particle numbers. Appropriate initial
conditions can in principle be used to obtain results for a specific
particle number. While this strategy tends to work well in real time
dynamics, the exponential increase of small wave function components
can results in unphysical results in imaginary time
propagation. Practical solutions to this problem are investigated in
this work.

The present article is organized as follows. After a brief review of
the \ac{MCTDH-oSQR} approach for real time propagation in section
\ref{SEC::realtime}, the equations of motions for the imaginary time
are discussed in Sec. \ref{SEC::EOM}. In Sec. \ref{SEC::gauge},
different gauges are discussed and a new gauge which can be
efficiently employed in imaginary time propagation is
introduced. Constraining operators which guarantee particle number
conservation during propagation in imaginary time are described in
Sec. \ref{SEC::conservation}. The different aspects are illustrated by
numerical examples studying Bose-Hubbard models in
Sec. \ref{SEC::example}. An important result of the numerical
study, the conservation of local particle numbers at all layers
of wavefunction representation, is discussed in more detail in
Sec. \ref{SEC::Discussion}. There also an efficient tensor contraction 
scheme utilizing the particle number conservation to reduce the 
numerical effort is suggested. Finally, concluding remarks are given 
in Sec. \ref{SEC::Conclusion}.

\section{MCTDH-\MakeLowercase{o}SQR}
\label{SEC::realtime}

In the \ac{MCTDH-oSQR} approach\cite{MW} the wave function is
represented in a basis of configurations,
\begin{align} 
&\left| \Psi \right> = \sum_{i_1} .. \sum_{i_f} A_{i_1.. i_f}(t)\,\left| \Phi_{i_1.. i_f}(t)\right>.
\end{align} 
These configurations are direct products of powers of $f$
time-dependent creation operators acting on the vacuum state:
\begin{align} 
&\left| \Phi_{i_1.. i_f}(t)\right> = \frac{\left(\hat b_{1}^\dagger(t)\right)^{i_1}}
{\sqrt{i_1!}}  \,..\,\frac{\left(\hat b_{f}^\dagger(t)\right)^{i_f}}{\sqrt{i_f!}} \left|0\right>.
\end{align} 
The $f$ time-dependent creation operators $\hat b^\dagger_\mu$ are represented as linear
combinations of $F$ time-independent creation operators $\hat a^\dagger_k$:
\begin{align} 
  \hat b^\dagger_\mu(t) = \sum_{k=1}^F c_{\mu k}(t)\,\hat a^\dagger_k.
\end{align} 
The coefficients $c_{\mu k}$ (called orbitals) satisfy the
orthonormality relation
\begin{align} 
\label{EQ::orthonormal}
&\sum_{k=1}^F c^*_{\mu k}(t) c_{\nu k}(t) = \delta_{\mu \nu}
\end{align} 
and thus obey
\begin{align} 
&\sum_{k=1}^F c^*_{\mu k}(t)\,\frac{d c_{\nu k}}{dt} 
 =-\sum_{k=1}^F \frac{dc^*_{\mu k}}{dt}\, c_{\nu k}(t).
\end{align} 

The equations of motions for $A_{i_1.. i_f}(t)$ and $c_{\mu k}(t)$ are
derived employing the Dirac Frenkel variational principle\cite{MW}
\begin{align}
  \left< \delta\Psi \middle| i \frac{d}{dt} - \hat H \middle| \Psi\right>.
\end{align}
The orthonormality of the coefficients $c_{\nu k}$ is enforced by the
constraint that
\begin{align}
  \label{EQ::gauge}
  h_{\mu \nu} = \sum_{k=1}^F c_{\mu k}^*\, i\, \frac{dc_{\nu k}}{dt}
\end{align}
is an hermitian matrix. This constraint corresponds to the gauge
condition in the usual \ac{MCTDH} formalism. Therefore the matrix
$h_{\mu \nu}$ will be called 'gauge' matrix despite the fact that it
does not correspond to a true gauge invariance in the present
context.\cite{MW}

The resulting equation of motion for the tensor $A_{i_1.. i_f}(t)$ is
\begin{align}
  \label{EQ::dAdt}
  &i\,\frac{dA_{i_1 .. i_f}}{dt} = \left< \Phi_{i_1 .. i_f} \middle| \hat H 
    - \sum_{\mu, \nu=1}^f\hat b^\dagger_\nu  h_{\nu \mu} \hat b_\mu \middle| \Psi\right>.
\end{align}
This equation shows an analogous structure to the standard
\ac{MCTDH} equation of motion for the coefficient tensor. 
The MCTDH-oSQR approach can thus straightforwardly be generalized to
a multi-layer MCTDH-oSQR scheme.  
The equations of motion of the orbitals $c_{\mu k}(t)$ read
\begin{align}
  \label{EQ::dcdt}
  &i\,\frac{dc_{\mu k}}{dt}
  = \sum_{\nu=1}^f  h_{\mu \nu} c_{\nu k} + \sum_{l=1}^F \left(\delta_{k l}
  - \sum_{\nu=1}^f c_{\nu k} c^*_{\nu l}\right) \nonumber\\&~~~~~
  ~~~~~~\cdot\sum_{\lambda=1}^f \rho^{-1}_{\mu \lambda}
    \left<\Psi\middle| \hat b_\lambda^\dagger
    \middle[\hat a_l, \hat H\middle]_{\hat a_i=\sum_{\kappa=1}^f c_{\kappa i} \hat b_\kappa}\middle|\Psi\right> 
\end{align}
where
\begin{align}
    \rho_{\mu \lambda} = \left<\Psi\middle| \hat b_\mu^\dagger \hat b_\lambda \middle|\Psi\right>.
\end{align}
The Hamiltonian $H$ appearing in Eq. (\ref{EQ::dcdt}) is typically
written employing the operators $a_i$ and $a_i^\dagger$ in normal
ordering to facilitate an efficient evaluation of the expectation
value $\left<\Psi\middle| \hat b_\lambda^\dagger\middle[\hat a_l,
  \hat H\middle]_{\hat a_i=\sum_{\kappa=1}^f c_{\kappa i} \hat b_\kappa}
  \middle|\Psi\right> $.  The subscript $\hat a_i=\sum_{\kappa=1}^f
c_{\kappa i} \hat b_\kappa$ indicates that all time-independent operators
$\hat a_i$ and $\hat a_i^\dagger$ are replaced by time-dependent
operators $\hat b_\kappa$ and $\hat b_\kappa^\dagger$ after evaluation of the
commutator $\left[\hat a_l,\hat H\right]$.

\section{Equations of Motion in imaginary time}
\label{SEC::EOM}

Considering \ac{MCTDH-oSQR} propagation in imaginary time, $\tau =
i\,t$, the ansatz for the wave function analogously reads
\begin{align} 
\label{EQ::ImagWF}
  &\left| \Psi \right> = \sum_{i_1} .. \sum_{i_f} A_{i_1.. i_f}(\tau)\,\left| \Phi_{i_1.. i_f}(\tau)\right>, \\
\label{EQ::ImagWF2}
  & \left| \Phi_{i_1.. i_f}(\tau)\right> = \frac{\left(\hat b_{1}^\dagger(\tau)\right)^{i_1}}
  {\sqrt{i_1!}}  \,..\,\frac{\left(\hat b_{f}^\dagger(\tau)\right)^{i_f}}{\sqrt{i_f!}} \left|0\right>, \\
\label{EQ::ImagWF3}
  &  \hat b^\dagger_\mu(\tau) = \sum_{k=1}^F c_{\mu k}(\tau)\,\hat a^\dagger_k.
\end{align} 
The orbitals $c_{\mu k}(\tau)$ obey orthonormality relations
analogous to Eq. (\ref{EQ::orthonormal}).

The equations of motion for imaginary time propagation can be derived
analogously to the ones for real time propagation (an explicit
derivation is given in the Appendix) and read
\begin{align}
\label{EQ::AEOMIM}
&-\,\frac{dA_{i_1 .. i_f}}{d\tau} = \left< \Phi_{i_1 .. i_f} \middle| \hat H 
- \sum_{\nu, \mu = 1}^f \hat b^\dagger_\nu \bar h_{\nu \mu} \hat b_\mu \middle| \Psi\right>,\\
\label{EQ::CEOMIM}
&-\frac{dc_{\mu k}}{d\tau}=
\sum_{\nu = 1}^f \bar h_{\mu \nu} c_{\nu k}
+  \sum_{l=1}^F \left(\delta_{k l} - \sum_{\nu=1}^f c_{\nu k} c^*_{\nu l} \right) 
\nonumber\\ & ~~~~~~~~
~~~~~~\cdot\sum_{\lambda=1}^f \rho^{-1}_{\mu \lambda}
\left<\Psi \middle| \hat b^\dagger_\lambda \left[\hat a_l,\hat H\right]_{\hat a_i=\sum\limits_{\kappa=1}^f c_{\kappa i} \hat b_\kappa}\middle| \Psi\right>.
\end{align}
The only difference between real and imaginary time propagation is a
change in the matrix resulting from the 'gauge' condition. The matrix
\begin{align}
\label{EQ::ImagGauge}
\bar h_{l \mu} = -\sum_{k=1}^F c^*_{l k} \frac{dc_{\mu k}}{d\tau}.
\end{align}
appearing in the imaginary time equations of motions is
anti-hermitian. The evolution of the time-dependent creation and
annihilation operators is then given by
\begin{align}
\label{EQ::dbdagger}
  -\frac{d\hat b_\mu^\dagger}{d\tau} &= -\sum_{k=1}^F \frac{dc_{\mu k}}{d\tau}\,\hat a_k^\dagger
  = -\sum_{k,l=1}^F c^*_{l k} \frac{dc_{\mu k}}{d\tau}\,\hat b_l^\dagger 
  = \sum_{l=1}^F \bar h_{l \mu} \hat b_l^\dagger,\\
%%%%%%%%%%%%%%%%
\label{EQ::db}
  -\frac{d\hat b_\mu}{d\tau} &= -\sum_{k=1}^F \frac{dc^*_{\mu k}}{d\tau}\,\hat a_k
  = -\sum_{k,l=1}^F c_{l k} \frac{dc^*_{\mu k}}{d\tau}\,\hat b_l
  = \sum_{l=1}^F \bar h^*_{l \mu} \hat b_l.
\end{align}

\section{'Gauge' Conditions}
\label{SEC::gauge}

The equations of motion Eq. (\ref{EQ::AEOMIM}) and (\ref{EQ::CEOMIM})
contain the 'gauge' matrix $\bar h_{\mu \nu}$ which introduces
rotations in the space spanned by the orbitals $c_{\nu k}$. For a
fixed number of basis functions the quality of the representation
depends on the orbital rotations.\cite{MW} Thus, the choice of the
'gauge' matrix $\bar h_{\mu\nu}$ affects the accuracy of a
\ac{MCTDH-oSQR} calculation for a given basis size.

The typical choice for the 'gauge' matrix $\bar h_{\mu \nu}$ used in
\ac{MCTDH} calculations is
\begin{align}
\bar h^{(zero)}_{\mu \nu} = 0.
\end{align}
This choice will be called \textit{zero gauge} in the following. The
zero gauge simplifies the equations of motions but does in general not
minimize the correlation in the second quantization representation. In
contrast, the use of a natural orbital representation minimizes the
apparent correlation in a two particle system\cite{L55NatOrb} and
approximately minimizes the correlation in a multi-particle system
(see, e.g., Ref. \onlinecite{JM2} for a discussion of the issue).  A natural
orbital representation, which implies a diagonal form of the
one-particle density matrix,
\begin{align}
\label{EQ::NatCon}
\rho_{\mu \nu}(\tau) = \delta_{\mu \nu}\,\rho_\nu(\tau),
\end{align}
can be retained during the propagation if a specific 'gauge' matrix
$\bar h_{\mu\nu}^{(nat)}$ is employed. This choice is called \textit{natural
gauge} in the following.

The natural gauge matrix $\bar h_{\mu\nu}^{(nat)}$ can be derived by
considering the time derivative of the one-particle density matrix:
\begin{align} 
  \frac{d\rho_{\mu \nu}}{d\tau}
  &= \frac{d}{d\tau} \left< \Psi \middle| \hat b^\dagger_\mu \hat b_\nu \middle| \Psi \right>
%%%%%%%%%%%%%%%%%%%
  \nonumber\\&=
  - \left< \Psi \middle| \hat H \hat b^\dagger_\mu \hat b_\nu \middle| \Psi \right>
  - \left< \Psi \middle| \hat b^\dagger_\mu \hat b_\nu \hat H \middle| \Psi \right>
  \nonumber\\&~~~~
  + \left< \Psi \middle| \frac{d \hat b^\dagger_\mu}{d\tau} \hat b_\nu \middle| \Psi \right>
  + \left< \Psi \middle| \hat b^\dagger_\mu \frac{d \hat b_\nu}{d\tau} \middle| \Psi \right>
%%%%%%%%%%%%%%%%%%%%
%  \nonumber\\&=
%  - \left< \Psi \middle| H b^\dagger_\mu b_\nu \middle| \Psi \right>
%  - \left< \Psi \middle| b^\dagger_\mu b_\nu H \middle| \Psi \right>
%  \nonumber\\&~~~~
%  + \sum_{k=1}^F\left< \Psi \middle| \frac{d c_{\mu k}}{d\tau} a_k^\dagger b_\nu \middle| \Psi \right>
%  + \sum_{k=1}^F\left< \Psi \middle| b^\dagger_\mu \frac{d c^*_{\nu k}}{d\tau} a_k \middle| \Psi \right>
%%%%%%%%%%%%%%%%%%%%
%  \nonumber\\&=
%  - \left< \Psi \middle| H b^\dagger_\mu b_\nu \middle| \Psi \right>
%  - \left< \Psi \middle| b^\dagger_\mu b_\nu H \middle| \Psi \right>
%  \nonumber\\&~~~~
%  + \sum_{\sigma=1}^f \sum_{k=1}^F c^*_{\sigma k} \frac{d c_{\mu k}}{d\tau} \left< \Psi \middle|  b_\sigma^\dagger b_\nu \middle| \Psi \right>
%  + \sum_{\sigma=1}^f \sum_{k=1}^F c_{\sigma k} \frac{d c^*_{\nu k}}{d\tau} \left< \Psi \middle| b^\dagger_\mu b_\sigma \middle| \Psi \right>
%%%%%%%%%%%%%%%%%%%%
  \nonumber\\&=
  - \left< \Psi \middle| \hat H \hat b^\dagger_\mu \hat b_\nu \middle| \Psi \right>
  - \left< \Psi \middle| \hat b^\dagger_\mu \hat b_\nu \hat H \middle| \Psi \right>
  \nonumber\\&~~~
  - \sum_{\sigma=1}^f \bar h_{\sigma \mu} \rho_{\sigma \nu}
  - \sum_{\sigma=1}^f \bar h^*_{\sigma \nu} \rho_{\mu \sigma}
\end{align}
Here the definition of the 'gauge' matrix from Eq. (\ref{EQ::ImagGauge})
was used. The derivative of the one particle density matrix can be
written in compact form employing the anti-commutator $\{A,B\} = A B +
B A$ and the anti-hermiticity of $\bar h_{\sigma\nu}$:
\begin{align} 
  \frac{d\rho_{\mu \nu}}{d\tau}&=
  - \left< \Psi \middle| \middle\{\hat H, \hat b^\dagger_\mu \hat b_\nu\middle\} \middle| \Psi \right> 
  - \sum_{\sigma=1}^f \left(\bar h_{\sigma \mu} \rho_{\sigma \nu} -\bar h_{\nu \sigma} \rho_{\mu \sigma}   \right).
\end{align} 
In the natural orbital representation 
\begin{align}  
0=\frac{d\rho_{\mu\nu}}{d\tau}
\end{align}  
and thus
\begin{align}  
\label{EQ::NATCONDITION}
0&=\left< \Psi \middle| \middle\{\hat H, \hat b^\dagger_\mu \hat b_\nu\middle\}
  \middle| \Psi \right>
+ \sum_{\sigma=1}^f \left(\bar h^{(nat)}_{\sigma \mu} \rho_{\sigma \nu} - \bar h^{(nat)}_{\nu \sigma} \rho_{\mu \sigma} \right)\nonumber\\
&=\left< \Psi \middle| \middle\{\hat H, \hat b^\dagger_\mu \hat b_\nu\middle\}
  \middle| \Psi \right> + \bar h^{(nat)}_{\nu\mu} \left(\rho_{\nu} -
  \rho_{\mu}\right)
\end{align}
has to be fulfilled for all $\mu\ne\nu$.  Employing a regularization
which guarantees that \mbox{$\bar h_{\mu \nu} \rightarrow 0$} for
\mbox{$\rho_{\mu} - \rho_{\nu} \rightarrow 0$}, one finally obtains
\begin{align} 
\label{EQ::IMNATGAUGE}
  \bar h^{(nat)}_{\nu \mu} = \left< \Psi \middle| \middle\{\hat H, \hat b^\dagger_\mu \hat b_\nu\middle\} \middle| \Psi \right>
  \frac{\rho_{\nu} - \rho_{\mu}}{(\rho_{\nu} - \rho_{\mu})^2 + \epsilon^2}
\end{align}
where $\epsilon$ is a small positive regularization constant.

It is interesting to compare the result for the imaginary time
propagation, Eq. (\ref{EQ::IMNATGAUGE}), with the one for the real-time
propagation:
\begin{align}
\label{EQ::RENATGAUGE}
h^{(nat)}_{\nu \mu} &= \left< \Psi \middle| \middle[\hat H, \hat b^\dagger_\mu \hat b_\nu\middle] \middle| \Psi \right>
\frac{\rho_{\nu} - \rho_{\mu}}{(\rho_{\nu} - \rho_{\mu})^2 + \epsilon^2}
\end{align}
(see Ref. \onlinecite{MW} for a derivation). At this point an important
difference between the anti-commutator $\{\hat H, \hat b^\dagger_\mu \hat b_\nu\}$ in
Eq. \ref{EQ::IMNATGAUGE} and the commutator $[\hat H, \hat b^\dagger_\mu \hat b_\nu]$
in Eq. \ref{EQ::RENATGAUGE} should be noted. Considering Hamiltonians
including up to N-particle operators, the commutator can include in
maximum N-particle operators while the anti-commutator can include
(N+1)-particle operators. This can be nicely illustrated for the
one-particle Hamiltonian
\begin{align}
  \hat H^{(1)} = \sum_{\rho,\sigma=1}^f w_{\rho \sigma} \hat b^\dagger_\rho \hat b_\sigma.
\end{align}
Here one finds
\begin{align}
  \{\hat H, \hat b^\dagger_\mu \hat b_\nu\}
  &= \sum_{i,j=1}^f w_{\rho \sigma} \hat b^\dagger_\rho \hat b_\sigma \hat b^\dagger_\mu \hat b_\nu 
  + \sum_{i,j=1}^f  w_{\rho \sigma} \hat b^\dagger_\mu \hat b_\nu \hat b^\dagger_\rho \hat b_\sigma
  \nonumber \\
  &= 2 \sum_{\rho,\sigma=1}^f w_{i j} \hat b^\dagger_\mu \hat b^\dagger_i \hat b_j \hat b_\nu 
  \nonumber\\&~~~
  + \sum_{\rho=1}^f w_{\rho \mu} \hat b^\dagger_\rho \hat b_\nu 
  + \sum_{\sigma=1}^f  w_{\nu \sigma} \hat b^\dagger_\mu \hat b_\sigma
\end{align}
and
\begin{align}
  [\hat H, \hat b^\dagger_\mu \hat b_\nu]
  &= \sum_{i,j=1}^f w_{\rho \sigma} \hat b^\dagger_\rho \hat b_\sigma \hat b^\dagger_\mu \hat b_\nu 
  - \sum_{i,j=1}^f  w_{\rho \sigma} \hat b^\dagger_\mu \hat b_\nu \hat b^\dagger_\rho \hat b_\sigma
  \nonumber \\
  &= \sum_{\rho=1}^f w_{\rho \mu} \hat b^\dagger_\rho \hat b_\nu 
  + \sum_{\sigma=1}^f  w_{\nu \sigma} \hat b^\dagger_\mu \hat b_\sigma.
\end{align}
This difference is of central importance for the numerical effort
involved in the imaginary time propagation. If the system of interest
is described by a N-Particle Hamiltonian and $f$ orbitals, the effort
to calculate $\left< \Psi \middle|\middle\{\hat H, \hat b^\dagger_\mu
  \hat b_\nu\middle\} \middle| \Psi \right>$ scales proportional to
$f^{2N+2}$. In contrast, any other object present in the
\ac{MCTDH-oSQR} approach can be evaluated with an effort scaling as
$f^{2N}$. Thus, the use of the natural gauge condition might reduce
correlation but it significantly increases the numerical effort for a
given number of orbitals.

Alternatively, a gauge reducing the spectral width of the
effective operator driving the motion of the wave function can be
considered. To this end, the measure
\begin{align}
\label{EQ::Measure}
\Delta &= \sum_\mu\left(\frac{d}{d\tau} \left<\hat b_\mu \Psi\right|\right)
\left(\frac{d}{d\tau} \left|\hat b_\mu \Psi\right>\right)
\end{align}
is minimized with respect to $h_{\mu \nu}$. The resulting 'gauge' will
be called \textit{spectral gauge} in the following. Employing
$\frac{d\Psi}{d\tau}=-\hat H\Psi$ and Eqs. (\ref{EQ::dbdagger}),
(\ref{EQ::db}), the measure $\Delta$ reads
\begin{align}
\Delta
&=\sum_{\mu = 1}^f\left<\hat b_\mu \hat H \Psi\middle| \hat b_\mu \hat H \Psi\right>
+ \sum_{\mu,\nu = 1}^f \left<\hat b_\mu \hat H \Psi\middle| \bar h_{\nu \mu}^* \hat b_\nu \Psi\right>
\nonumber\\&
~~~+ \sum_{\mu,\nu = 1}^f \left<\bar h_{\nu \mu}^* \hat b_\nu \Psi\middle| \hat b_\mu \hat H \Psi\right>
+ \sum_{\mu,\nu,\lambda = 1}^f \left<\bar h_{\lambda \mu}^* \hat b_\lambda\Psi\middle| \bar h_{\nu \mu}^* \hat b_\nu \Psi\right>
\nonumber\\%%%%%%%%%%%%%%%%%%
&=\sum_{\mu = 1}^f\left<\hat b_\mu \hat H \Psi\middle| \hat b_\mu \hat H \Psi\right>
+ \sum_{\mu,\nu = 1}^f \bar h_{\nu \mu}^* \left<\hat b_\mu \hat H \Psi\middle| \hat b_\nu \Psi\right>
\nonumber\\ &
~~~+ \sum_{\mu,\nu = 1}^f \bar h_{\nu \mu} \left<\hat b_\nu \Psi\middle| \hat b_\mu \hat H \Psi\right>
+ \sum_{\mu,\nu,\lambda = 1}^f \bar h_{\lambda \mu} \bar h_{\nu \mu}^*\left<\hat b_\lambda\Psi\middle| \hat b_\nu \Psi\right>.
\end{align}
If the anti-hermiticity of the $\bar h$-matrix is employed, one finds
\begin{align}
\Delta&=\sum_{\mu = 1}^f\left<\hat b_\mu \hat H \Psi\middle| \hat b_\mu \hat H \Psi\right>
- \sum_{\mu,\nu = 1}^f  \bar h_{\mu \nu}\left<\hat b_\mu \hat H \Psi\middle| \hat b_\nu \Psi\right>
\nonumber\\ &
~~~+ \sum_{\mu,\nu = 1}^f \bar h_{\nu \mu} \left<\hat b_\nu \Psi\middle| \hat b_\mu \hat H \Psi\right>
- \sum_{\mu,\nu,\lambda = 1}^f \bar h_{\lambda \mu}\bar h_{\mu \nu}\left<\hat b_\lambda\Psi\middle| \hat b_\nu \Psi\right>.
\end{align}
Minimizing $\Delta$ with respect to $\bar h_{\sigma \xi}$ yields
\begin{align}
0 &= \frac{\partial \Delta}{\partial \bar h_{\sigma \xi}} 
\nonumber\\
&= -\left< \Psi \middle| \hat H \hat b^\dagger_\sigma \hat b_\xi \middle| \Psi \right>
+ \left<\Psi\middle|\hat b^\dagger_\sigma \hat b_\xi \hat H \middle|\Psi\right>
\nonumber\\ &
~~~- \sum_{\lambda = 1}^f \bar h_{\lambda \sigma} \left<\hat b_\lambda\Psi\middle| \hat b_\xi \Psi\right>
- \sum_{\nu = 1}^f \bar h_{\xi \nu}\left<\hat b_\sigma\Psi\middle| \hat b_\nu \Psi\right> 
\nonumber\\
&= - \left< \Psi \middle| \left[\hat H, \hat b^\dagger_\sigma \hat b_\xi\right] \middle| \Psi \right>
- \sum_{\lambda = 1}^f \bar h_{ \lambda \sigma} \rho_{\lambda \xi}
- \sum_{\lambda = 1}^f  \bar h_{\xi \lambda} \rho_{\sigma \lambda}.
\end{align}
Thus, the $\bar h^{(spec)}$-matrix corresponding to the spectral gauge
has to fulfill:
\begin{align}
\label{EQ::ToSolve}
\sum_{\lambda = 1}^f \left(\bar h^{(spec)}_{\lambda \sigma} \rho_{\lambda \xi}
+  \bar h^{(spec)}_{\xi \lambda} \rho_{\sigma \lambda}  \right)
&= - \left<\Psi\middle|\left[\hat H, \hat b^\dagger_\sigma \hat b_\xi\right] \middle|\Psi\right>.
\end{align}
To obtain an explicit expression for $\bar h^{(spec)}$, one first
transforms Eq. (\ref{EQ::ToSolve}) into the natural orbital
representation.  The matrix $u$ transforming from any given
representation into the natural orbital representation can be obtained
by diagonalizing the single-particle density matrix:
\begin{align}
\label{EQ::NatRepa}
\tilde \rho_{\mu} = \sum_{\delta,\lambda=1}^f u^*_{\lambda \mu} \,\rho_{\lambda \delta}\, u_{\delta \mu}.
\end{align}
The transformed matrices and operators read
\begin{align}
\label{EQ::NatRepb}
\tilde h_{\nu \mu} &= \sum_{\delta,\lambda=1}^f u^*_{\lambda \mu} \bar h_{\delta \lambda} u_{\delta \nu}.\\
\label{EQ::NatRepc}
\hat{\tilde b}^\dagger_\mu &= \sum_{\delta,\lambda=1}^f u^*_{\lambda \mu} \hat b^\dagger_\lambda\\
\label{EQ::NatRepd}
\hat{\tilde b}_\mu &= \sum_{\delta,\lambda=1}^f u_{\lambda \mu} \hat b_\lambda
\end{align}
Eq.~(\ref{EQ::ToSolve}) then takes the simplified form
\begin{align}
\tilde h^{(spec)}_{\xi \sigma} \left(\tilde \rho_{\xi}
+ \tilde \rho_{\sigma}\right)
&= \left<\Psi\middle|\left[\hat H, \hat{\tilde b}^\dagger_\lambda \hat{\tilde b}_\delta\right] \middle|\Psi\right>
\end{align}
Dividing by the regularized sum $\tilde \rho_{\rho}+ \tilde
\rho_{\sigma}+ \epsilon$ yields an explicit expression for the
spectral gauge matrix  $\tilde h^{(spec)}$ in natural representation:
\begin{align}
  \tilde h^{(spec)}_{\xi \sigma} &=
  \frac{\left<\Psi\middle|\left[\hat H, \hat{\tilde{b}}^\dagger_\sigma \hat{\tilde
        b}_\xi\right] \middle|\Psi\right>} {\tilde
    \rho_{\xi}+ \tilde \rho_{\sigma} + \epsilon}.
\end{align}
Transformation to the original representation results in the working
equation defining $\bar h_{\mu \nu}$ in the spectral gauge:
\begin{align}
  \label{EQ::SpecGauge}
\bar h^{(spec)}_{\mu \nu} 
&= \sum_{\xi,\sigma,\delta,\lambda=1}^f u_{\nu\sigma} \frac{u^*_{\lambda \sigma} 
\left<\Psi\middle|\left[\hat H, \hat b^\dagger_\lambda \hat b_\delta\right] \middle|\Psi\right>u_{\delta \xi}
}
{\tilde \rho_{\xi}+ \tilde \rho_{\sigma} + \epsilon} u^*_{\mu \xi}.
\end{align}
Since this equation contains a commutator and not a anti-commutator,
the numerical effort for the calculation of this so called spectral
gauge scales proportional to $f^{2N}$ if the solution of an
$N$-Particle Hamiltonian is described by $f$ orbitals. Moreover, the
numerical results shown in Sec. \ref{SEC::example} indicate that the
spectral gauge reduces the correlation in the wave function.  A simple
example that provides further insights into the advantages of the
spectral gauge is discussed in Appendix B.

\section{Particle Number Conservation}
\label{SEC::conservation}

In systems with particle number conserving Hamiltonians one is
frequently interested in obtaining solutions with a fixed number of
particles. The \ac{MCTDH-SQR} and \ac{MCTDH-oSQR} approaches employs
the Fock space and thus contains states with all possible particle
numbers.  In principle, this does not constitute a problem for a
numerically exact approach. In particle number conserving systems
the particle number is a constant of motion and the particle number
can be defined via the initial conditions chosen.  However, in
practical calculations numerical errors can result in small
populations of states with incorrect particle numbers. This typically
is not an important issue in real time propagations. Here small
contributions to the wave function with incorrect particle numbers do
not grow during the propagation for a particle number conserving
Hamiltonians.  In contrast, these numerical errors can cause serious
problems in imaginary time propagation. Here small contributions to
the wave function with incorrect particle numbers can grow
exponentially during the propagation if these contribution shows
energies below the energies of relevant states with the correct
particle number. In this situation, particle number conservation must
be explicitly enforced to guarantee numerical stability.

Particle number conservation can be assured via a penalty term $\hat
H_{con}$ added to the system's Hamiltonian $\hat H_{sys}$. $\hat
H_{con}$ artificially increase the energy of states with particle
numbers differing from the desired particle number $N$.  Thus, the
effective Hamiltonian
\begin{align}
\hat H=\hat H_{Sys}+\hat H_{con} 
\end{align}
is used in the propagation.  $\hat H_{con}$ should not affect states
showing the desired particle number $N$ and strongly increase the
energy of all states showing particle numbers deviating from $N$. A
straightforward choice of $\hat H_{con}$ would be
\begin{align}
\hat H_{con}&=\Gamma\cdot\left(\sum_{k=1}^F \hat a_k^\dagger \hat a_k-N\right)^2
\label{EQ::badPun}
\end{align}
where $\Gamma$ is a large positive number. Within the context of the
present work, this operator has one significant disadvantage. In
\ac{MCTDH-SQR} and \ac{MCTDH-oSQR} calculations the Hamiltonian is
typically expanded in a sum of products of single-particle
operators. In the required form, $\hat H_{con}$ reads
\begin{align}
\hat H_{con}&=\Gamma\cdot\sum_{k=1}^F\sum_{l=1}^F\hat a_k^\dagger \hat a_k\hat a_l^\dagger \hat a_l-2 \Gamma\cdot N\cdot\sum_{k=1}^F \hat a_k^\dagger \hat a_k+N^2.
\end{align}
Thus, the addition of $\hat H_{con}$ gives rise to $F^2+F+1$ additional
terms in the Hamiltonian.

A more efficient choice of $\hat H_{con}$ is\cite{MasterWeike}
\begin{align}
\hat H_{con}&=
\frac{2\Gamma}{\zeta^2}\cdot \cosh\left(\zeta\left(\sum_{k=1}^F \hat a_k^\dagger \hat a_k-N\right)\right)-\frac{2\Gamma}{\zeta^2}
\label{EQ::goodPun}
\end{align}
where $\Gamma$ is a large positive number and $\zeta$ is a small positive
number. A Taylor expansion of this operator in $\zeta$,
\begin{align}
\hat H_{con}&\approx \Gamma \cdot \left(\sum_{k=1}^F \hat a_k^\dagger \hat a_k-N\right)^2\nonumber\\
&~~~+\frac{\zeta^2\,\Gamma}{12}\cdot\left(\sum_{k=1}^F \hat a_k^\dagger \hat a_k-N\right)^4 ~,
\end{align}
demonstrates that this operator and the $\hat H_{con}$ of
Eq. (\ref{EQ::badPun}) are asymptotically equivalent for small
$\zeta$. Rewriting Eq. (\ref{EQ::goodPun}) in sum of products form
results in
\begin{align}
\hat H_{con}
&= \frac{2\Gamma}{\zeta^2} \left(e^{\zeta N} \prod_{k=1}^F e^{\zeta\,\hat a_k^\dagger \hat a_k}
  + e^{-\zeta N} \prod_{k=1}^F e^{-\zeta\,\hat a_k^\dagger \hat a_k}\right) - \frac{2\Gamma}{\zeta^2}.
\end{align}
This operator gives rise to only three additional terms in the
Hamiltonian.  Since a non-trivial $\hat H_{sys}$ consists of more than $F$
terms, the additional effort resulting from the application of this
$\hat H_{con}$ is small compared to numerical effort required for the
application of $\hat H_{sys}$.

\section{Numerical Examples}
\label{SEC::example}

\subsection{Implementation and Numerical Details}

Quantum dynamics calculations studying different Bose Hubbard models are 
used to illustrate different aspects of the methodology discussed above. 
ML-MCTDH-SQR and ML-MCTDH-oSQR propagation in imaginary time propagation 
is employed to calculate the respective ground states. A constant mean
field (CMF) approach \cite{MCTDHIntegrator,M4} is used to integrate the
ML-MCTDH-SQR and ML-MCTDH-oSQR equations of motion.

In the ML-MCTDH-SQR calculations the CMF2 scheme of Ref. \onlinecite{M4}
is employed. Since results obtained with different integration accuracies
will be compared in the context of the particle number conservation, a 
detailed description of the different relevant parameters will be given in
the following. The CMF approach employs a two level integration scheme with
different time steps used to update matrix elements and to integrate the
equations of motion for the expansion coefficients and the single-particle 
functions. Here a common accuracy parameter $\epsilon$ is used to control
the time steps at all levels: an accuracy parameter of $10 \cdot \epsilon$ 
is used in the CMF update step (see Eq.(27) of Ref. \onlinecite{M4} for 
the definition of the error estimate) and an accuracy 
parameter $\epsilon$ is employed in the Bulirsch-Stoer
integrator used to integrate the individual equations of motion (see 
Ref. \onlinecite{NumRec} for the definition of the error measure). Furthermore,
the same $\epsilon$ is used regularization constant in the computation
of the inverse density matrices.

\label{Example1}
\begin{figure}[t!]
\includegraphics[width=0.23\textwidth]{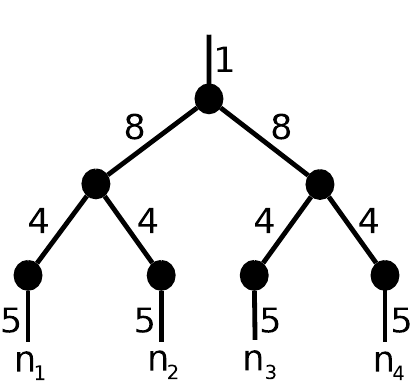}
\caption{\label{FIG::Tree-Conservation} Diagrammatic representation of
  \acs{ML-MCTDH-SQR} wave function used in Sect. \ref{Example1}.
  See Refs. \onlinecite{M6,MW} for a detailed description of the diagrammatic 
  notation. The occupation number $n_i$ of the i-th site takes the
  role of the i-th coordinate in the MCTDH-SQR approach.}
\end{figure} 

In the ML-MCTDH-oSQR approach the optimized orbitals $c_{\mu,k}(t)$
have to be propagated in addition to the expansion coefficients and 
single-particle functions of the ML-MCTDH ansatz already presented 
in the ML-MCTDH-SQR approach. Here an extended version of the CMF2 scheme
is used which treats the MCTDH-oSQR orbitals analogously to the 
single-particle functions (SPFs) in 
the ML-MCTDH approach. The 'gauge' matrix $\bar h_{\mu\nu}$,
the one particle density matrix
\begin{align}
  \left<\Psi\middle| \hat b^\dagger_\sigma \hat b_\xi\middle|\Psi\right>~,
\end{align}
and the two particle density matrix
\begin{align}
  \left<\Psi\middle| \hat b^\dagger_\lambda \hat b_\nu^\dagger \hat b_\kappa \hat b_\mu 
|\Psi\right>~
\end{align}
are only computed at CMF update steps. All other objects appearing in
the equations of motion of the optimized orbitals $c_{\mu,k}(t)$ can
be written as functions of the one and two particle density matrices
and the time-dependent orbitals $c_{\mu,k}(t)$. The integration of the
orbital equations of motion in the CMF approach then considers the
full time-dependence of the orbitals explicitly but assumes the gauge,
one particle density, and two particle density matrix elements do not
change during a CMF integration period. A Bulirsch-Stoer scheme is
employed for this integration. The common regularization parameter
$\epsilon$ is used in the implementation of Eqs. (\ref{EQ::CEOMIM}),
(\ref{EQ::IMNATGAUGE}) and (\ref{EQ::SpecGauge}).

\subsection{Particle Number Conservation}

First, the particle number conservation in the ML-MCTDH-SQR approach and 
the effect of the constraint operator $\hat H_{con}$ given in Eq.(\ref{EQ::goodPun})
is studied. Here a Bose Hubbard model described by the Hamiltonian
\begin{align}
  \label{Eq::Hubbard}
  \hat H_{Sys} =& -J~\sum_{i=1}^{F} \left(\hat a_i^\dagger \hat a_{i+1} + \hat a_{i+1}^\dagger \hat a_{i}\right)
  + \frac12 U \sum_{i=1}^{F} \left(\hat a_i^\dagger\right)^2 \hat a_{i}^2
\end{align}
is used. Twenty sides, $F=20$ and four particles, $N=4$, are
considered and the unity values of the parameters $J$ and $U$ are
assumed, $J=U=1$. Starting from the initial state
\begin{equation}
\left|\Psi(0)\right>= \hat a_1^\dagger \hat a_2^\dagger \hat a_3^\dagger \hat a_4^\dagger \left|0\right>,
\end{equation} 
the ground state is computed by imaginary time propagation. The
\ac{ML-MCTDH-SQR} wave function representation is given in
Fig. (\ref{FIG::Tree-Conservation}) (see Refs. \onlinecite{M6,MW} for
a detailed description of the diagrammatic notation). This representation
assures convergence with respect to the number of single-particle
functions employed.

\begin{figure}[t!]
{\centering\includegraphics[width=0.48\textwidth]{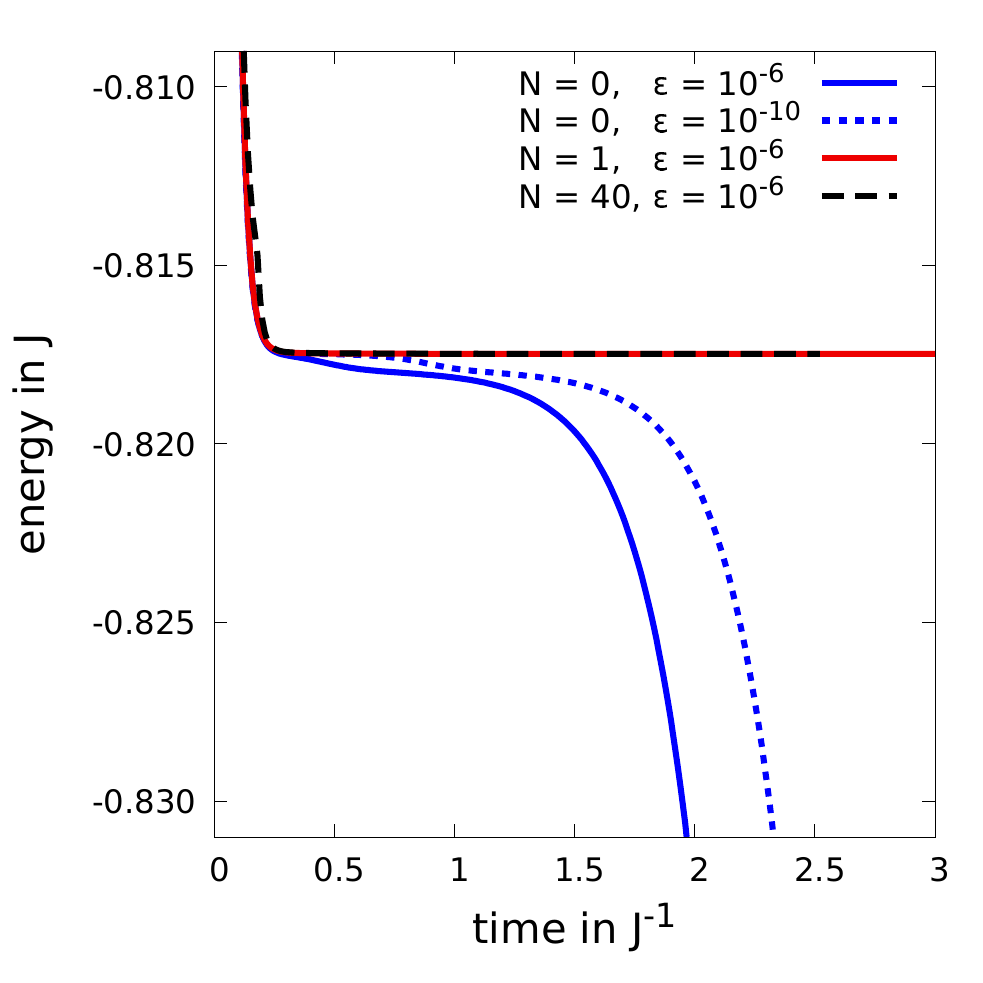}}
\caption{\label{ABB::ConservationZoom} The energy is displayed as a function of 
  the (imaginary) propagation time for different \acs{ML-MCTDH-SQR} 
  calculations. Results
  obtained with ($\Gamma=1$,$\Gamma=40$) and without ($\Gamma=0$) a constraint operator ensuring
  particle number conservation are displayed. Red and dashed black lines display
  results computed with constraint operators with $\Gamma=1$ and $\Gamma=40$, respectively.
  Solid and dotted blue lines show results of unconstrained calculations obtained
  with different integration accuracies ($\epsilon=10^{-6}$ and $\epsilon=10^{-10}$,
  respectively).}
\end{figure}

\begin{figure}[t!]
    \includegraphics[width=0.48\textwidth,trim=0 0 0 7]{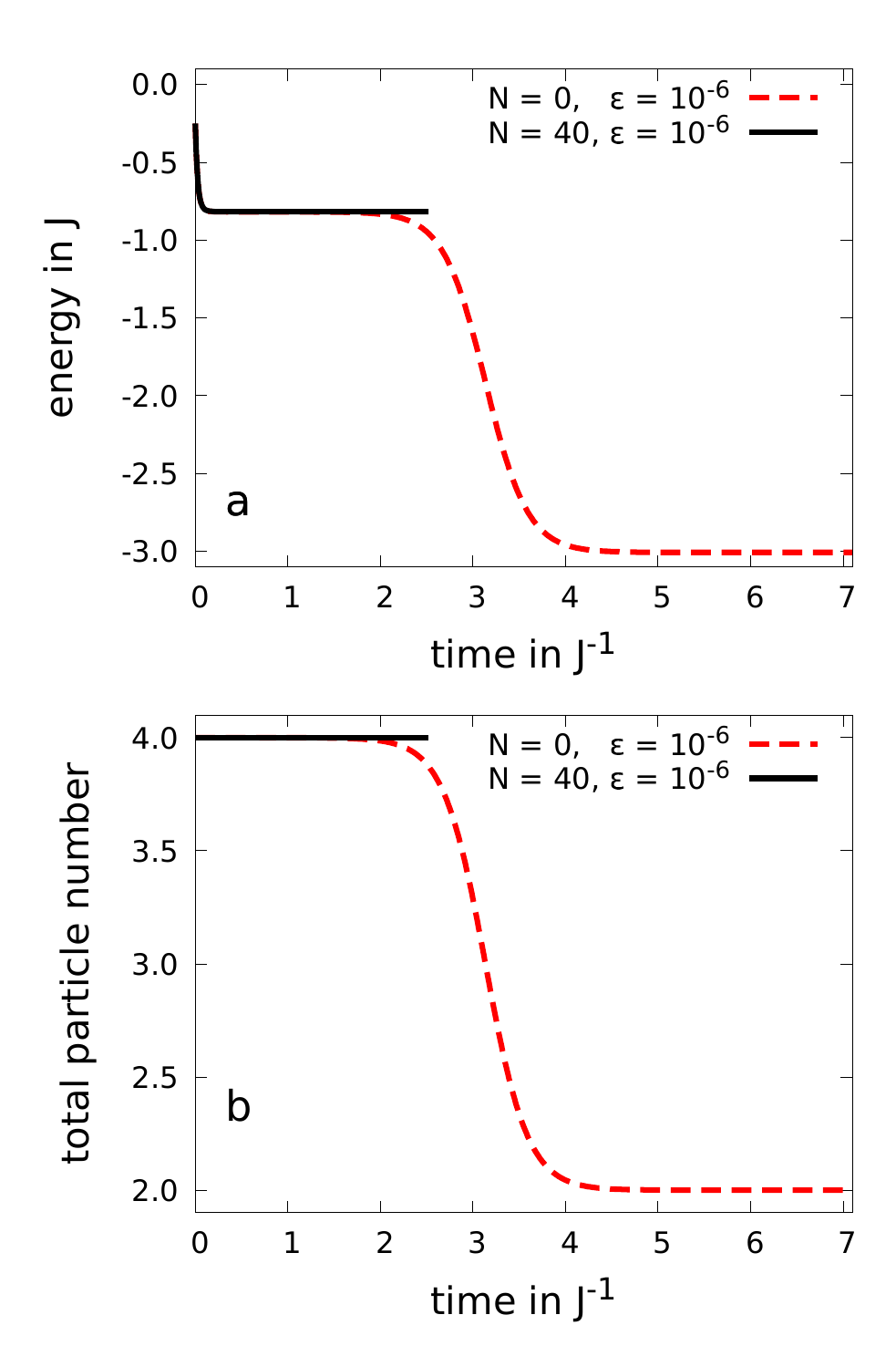}
    \caption{\label{ABB::Conservation} The expectation values of energy 
      and particle number are displayed as functions 
      of imaginary time in panels (a) and (b), respectively. 
      The black lines display the
      result of a calculation employing a constraint operator ensuring particle number
      conservation ($\Gamma=40$). The dashed red line shows the
      result obtained without constraint operator ($\Gamma=0$).}
\end{figure}

In Fig. \ref{ABB::ConservationZoom}, the energy relaxation during
the imaginary time propagation obtained from different calculations
is shown. Results obtained by straightforward calculations without constraint 
operator are given by blue lines. Results computed using the constraint operator 
specified in Eq.(\ref{EQ::goodPun}) with parameter values $\zeta=1/16$, $\Gamma=1$ or 
$\zeta=1/16$, $\Gamma=40$ are displayed by red or dashed black lines, respectively.
One immediately finds that the constrained calculations quickly converge to a ground 
state energy $E_0=0.817\,J$. This result is independent of the specific choice 
of the parameter $\Gamma$. In the unconstrained calculations, the energy
initially decays towards $E_0$, then stabilizes at values close to $E_0$ for an 
intermediate time period, and finally decays towards unphysically low values.

\begin{figure}[hbt!]
\centering\includegraphics[width=0.5\textwidth]{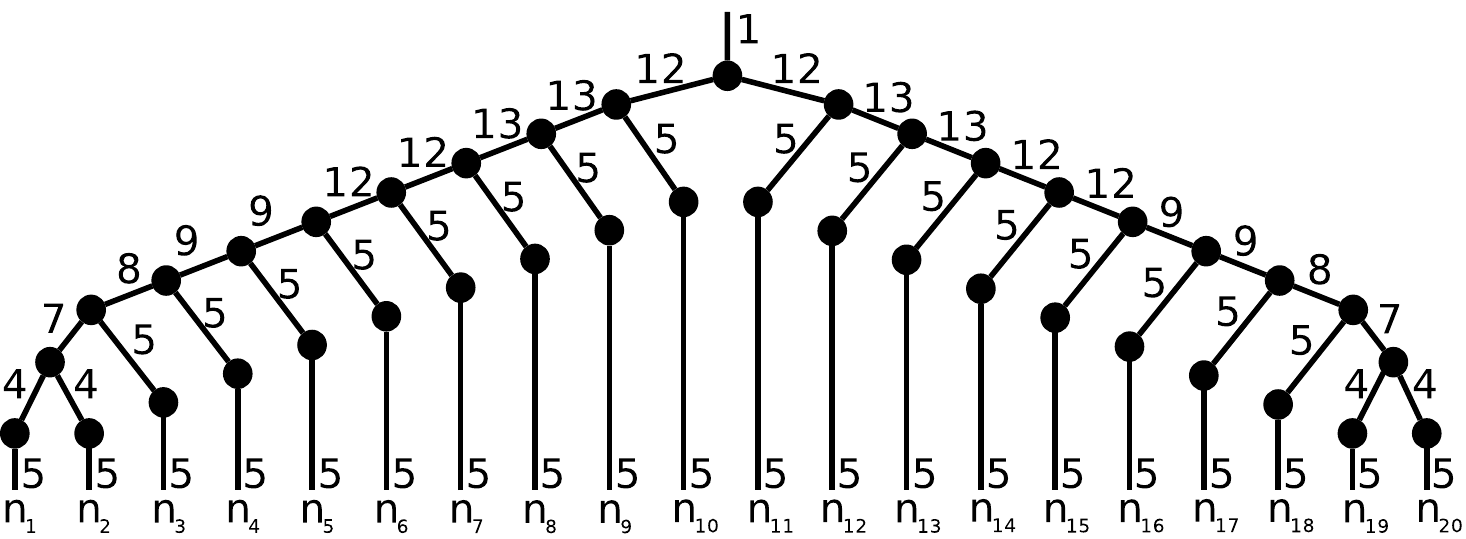}
\caption{\label{Abb::SQRWF} Graphical representation of the wavefunction
  used in the converged ML-MCTDH-SQR wavefunction reference calculations 
  discussed in Sect.\ref{Example2}.}
\end{figure}
Furthermore, the results of the constrained calculations crucially depend
on the integration accuracy: the solid blue line 
in Fig. \ref{ABB::ConservationZoom} shows the result obtained with the standard
integration accuracy $\epsilon=10^{-6}$ while the result of a high accuracy 
integration with $\epsilon=10^{-10}$ is displayed as dotted blue line. 
The more accurate calculation clearly shows a more extended intermediate plateau 
at energies close to $E_0$. This finding confirms the expectation that
a perfectly accurate integration should converge to the correct values even 
without the use a constraint operator.   

The connection between particle number conservation and numerical stability
is demonstrated in Fig. \ref{ABB::Conservation}. Comparing results obtained
with and without use of the constraint operator (displayed by black solid and
red dashed lines, respectively), panel (a) displays the expectation value of
energy and panel (b) shows the expectation values of the particle number as a 
function of the (imaginary) propagation time. One clearly sees that the
unphysical decay of the expectation value of energy coincides with the 
breakdown of the particle number conservation. The unconstrained calculation
finally converges towards a final state which shows a particle number of 
two and an energy of $-3 J$. The presence of a two-particle state with a 
lower energy in the Fock space thus causes the instability of the 
unconstrained \acs{ML-MCTDH-SQR} calculation. 

\subsection{MCTDH-oSQR in Imaginary Time}
\label{Example2}

The second numerical example studies single-layer and multi-layer
MCTDH-oSQR propagation in imaginary time for a particle number conserving 
Hamiltonian. Here a Bose-Hubbard model with
$N=4$ bosons on $F=20$ sides described by the Hamiltonian 
\begin{align}
  \label{Eq::Hubbard}
  \hat H =& -J~\sum_{i=1}^{F} \left(\hat a_i^\dagger \hat a_{i+1} + \hat a_{i+1}^\dagger \hat a_{i}\right)
  + \frac{1}{2} U \sum_{i=1}^{F} \left(\hat a_i^\dagger\right)^2 \hat a_{i}^2
  \nonumber\\
  &+ \sum_{i=1}^{F} \frac12\,\omega^2 \left(i-\frac{F+1}2\right)^2\, \hat a_i^\dagger \hat a_{i+1}.
\end{align}
with the parameter values $J=1$, $U=1$, $\omega=0.1$  is studied. 
This model has previously been used 
to illustrate the MCTDHB approach \cite{MCTDHBHubbard} is employed. 
Particle number conservation is explicitly enforced by the penalty operator $\hat H_{con}$ 
given in Eq. (\ref{EQ::goodPun}) with $\Gamma=500$ and $\zeta=0.05$.

\begin{figure}[hbt!]
\centering\includegraphics[width=0.45\textwidth]{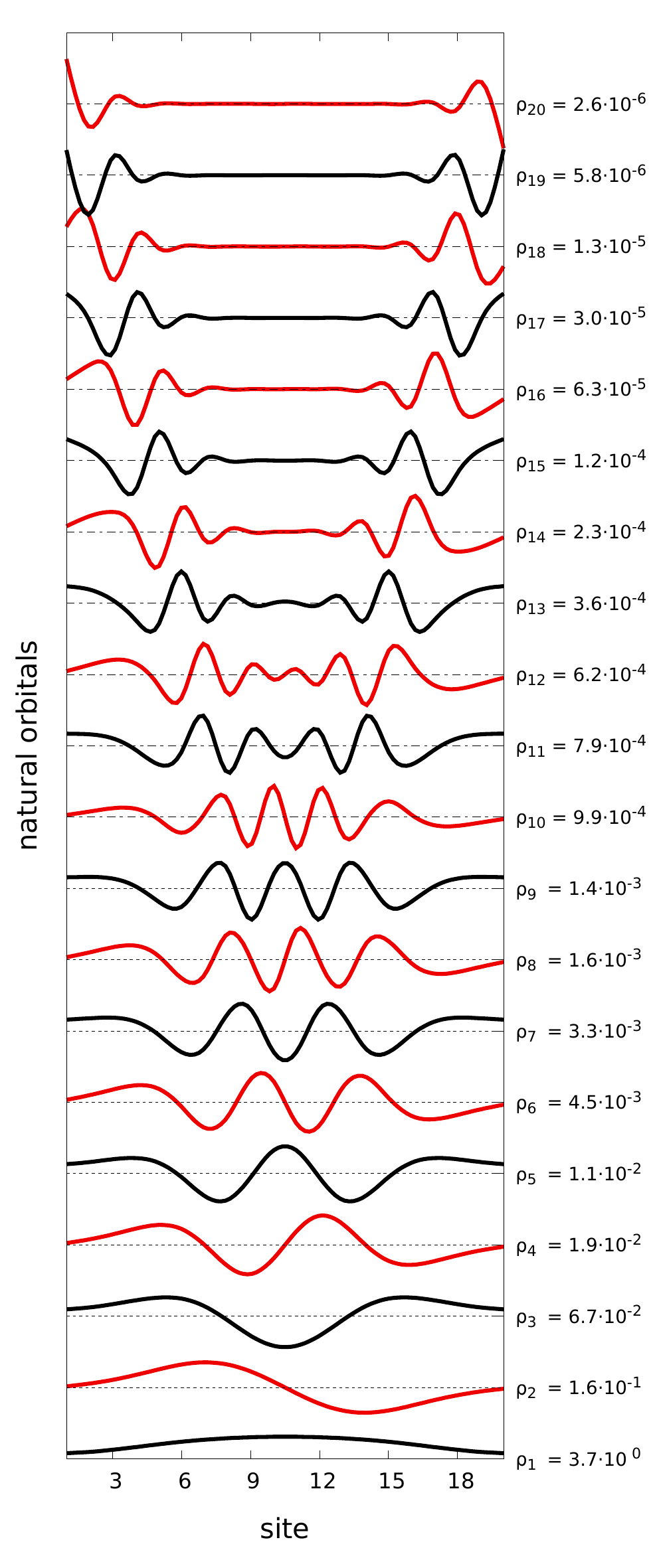}
\caption{\label{Abb::NatOrbSQR} Natural orbitals obtained from the
  reference calculation described in Sec. \ref{Example2}: the
  amplitude $c_{\mu k}$ of the natural orbital $\mu$ is displayed as
  a function of the side index $k$. The occupation number $n_\mu$ of
  each natural orbital $\mu$ is given.}
\end{figure}
A converged reference solution was calculated using the ML-MCTDH-SQR
approach. The diagrammatic representation of the wave function employed
is given in Fig. \ref{Abb::SQRWF}. A ground state energy of 
$E^{FCI}_{orb=20}=-7.3602(4)\,J$ is obtained. Natural orbitals are computed
by diagonalizing the one particle density matrix
\begin{align}
\rho_{i j} = \left<\Psi\middle|\hat a^\dagger_i \hat a_j\middle|\Psi\right>
\end{align}
for the ground state $\Psi$. In Fig. \ref{Abb::NatOrbSQR}, the
amplitudes $c_{\mu k}$ of the natural orbitals are plotted and 
the natural populations $\rho_\mu$ are given. A dominant fraction of the 
total population is found in a single natural orbital: the population
of the first natural orbital, $\rho_1=3.7$, accounts for 93\,\% of the 
total particle number of $4$. The population of the second most
populated natural orbital is more than a factor of $20$ smaller. 
The populations of the further natural orbitals then decrease roughly
exponentially. However, the decay is rather slow following roughly 
a $\rho_{\mu-1}/\rho_{\mu} \approx 2$ rule. Based on these results,
one can expect that a configuration interaction (CI) expansion in optimized 
orbitals yields reasonably accurate results with small numbers of 
orbitals but requires large number of orbitals to achieve high accuracies.
These expectations will be confirmed by the results of the \ac{MCTDH-oSQR}
calculations presented in the following.

\begin{figure}[b!]
\centering\includegraphics[width=0.25\textwidth]{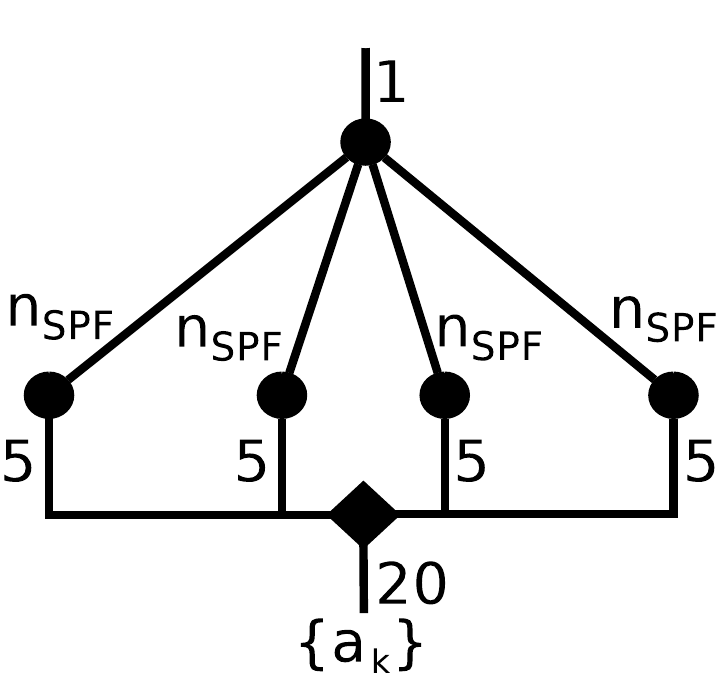}
\caption{\label{Abb::Gauge-Tree} Diagrammatic representation of MCTDH-oSQR
  wavefunction discussed in Sect.\ref{Example2}.}
\end{figure} 
The investigation of the imaginary time propagation in the \ac{MCTDH-oSQR} 
approach will first address the issue of the 'gauge' dependence discussed
in Sect.\ref{SEC::gauge}. To this end, \ac{MCTDH-oSQR} calculations
employing the zero gauge, the natural gauge, and the spectral gauge
are compared. The calculations employ $4$ optimized orbitals. The initial 
state at the start of the imaginary time propagation, $\Psi(0)$, shows
all 4 particles in a single orbital, i.e., 
\begin{equation}
\left|\Psi\right>=\frac1{\sqrt{4!}} \left(\hat b^\dagger_1\right)^4 \left|0\right> ~.
\end{equation}
The initial occupied orbital is the symmetric linear combination
of all $F$ sides, $c_{1 k} = 1/\sqrt{F},~k=1,..,F$. The real Bloch states
\begin{align}
c_{2 k} &= \frac1{\sqrt{2 F}} \sin(2\pi (k-0.5)/F), \\
c_{3 k} &= \frac1{\sqrt{2 F}} \cos(2\pi (k-0.5)/F), \\
c_{4 k} &= \frac1{\sqrt{2 F}} \sin(4\pi (k-0.5)/F) 
\end{align}
are chosen as the three initially unoccupied orbitals in the
\ac{MCTDH-oSQR} calculation. The unoccupied SPFs are determined by the scheme
described in Refs.~\onlinecite{M8,M10}.
\begin{figure}[t!]
\centering\includegraphics[width=0.45\textwidth]{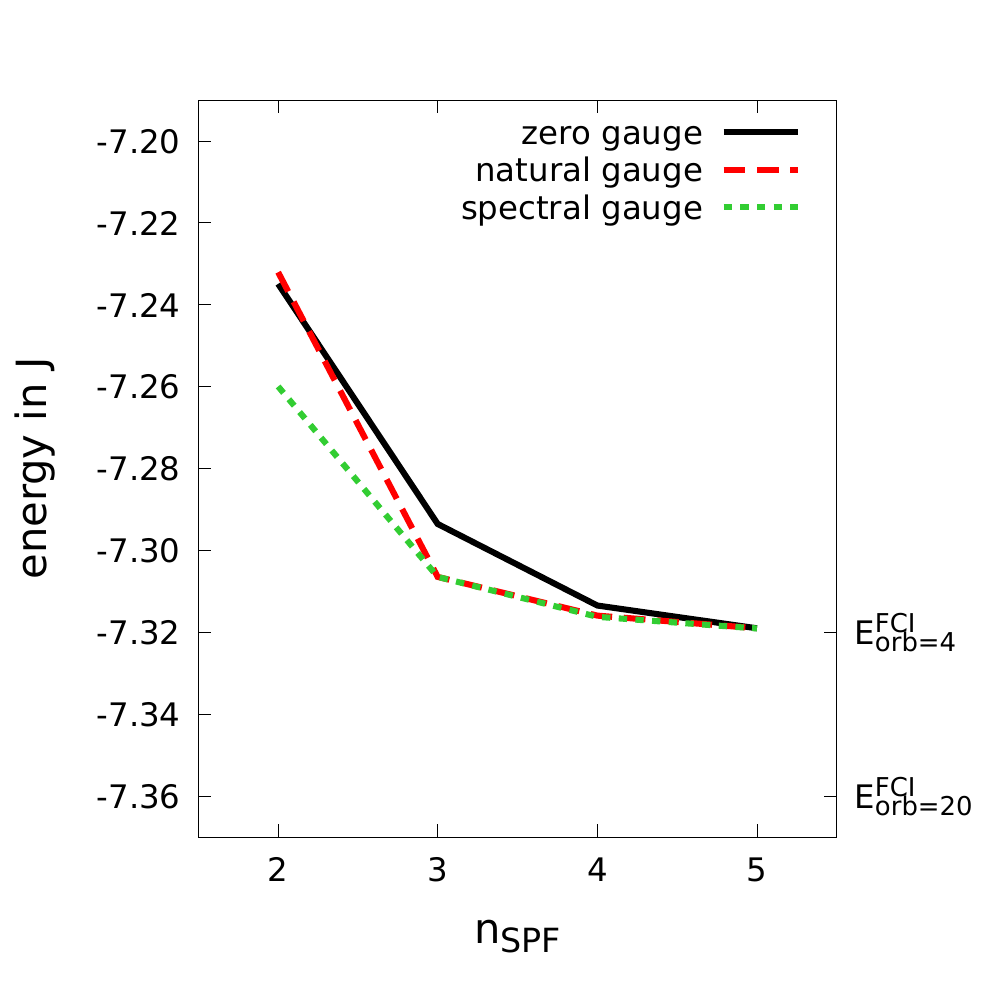}
\caption{\label{Abb::Gauge} Comparison of MCTDH-oSQR results obtained
  with zero gauge (solid black line), natural gauge (dashed red line),
  and spectral gauge (dotted green line), using the wavefunction
  representation depicted in Fig. \ref{Abb::Gauge-Tree}. The computed
  ground state energies are displayed as a function of the number of
  SPFs, $n_{SPF}$. The energy $E^{FCI}_{orb=20}$ computed by an exact
  ML-MCTDH-SQR calculations, and the energy $E^{FCI}_{orb=4}$, the
  MCTDH-oSQR result with four active orbitals converged with respect
  to the number of SPFs, are indicated.}
\end{figure}

Results of calculations employing the \ac{MCTDH-oSQR} representation
diagrammatically depicted in Fig. \ref{Abb::Gauge-Tree} with different
numbers of SPFs $n_{SPF}$ are shown in Fig. \ref{Abb::Gauge}.  The
computed ground state energy is displayed as a function of $n_{SPF}$
for calculations employing the zero gauge, the natural gauge, and the
spectral gauge. One finds that the optimal results for the ground
state energy which can be obtained with 4 active orbitals is
$E^{FCI}_{orb=4}=-7.32\,J$. This energy is $0.04\,J$ larger then the
accurate ground state energy $E^{FCI}_{orb=20}=-7.36 \,J$ obtained by
the reference calculations. Comparing the results obtained with
different 'gauges', differences in the convergence with respect to the
number of SPFs can seen in Fig. \ref{Abb::Gauge}. Calculations
employing the zero gauge converge more slowly than calculations
employing the natural gauge or the spectral gauge. Results obtained
with the natural gauge and the spectral gauge are essentially
identical for $n_{SPF}>2$. Surprisingly, for $n_{SPF}=2$ the spectral
gauge provides significantly better results than the natural gauge.
This result is presumably an artifact of the natural gauge calculation
resulting from incomplete convergence in the imaginary time propagation (see
below for a more detailed discussion). Considering the difference in
the numerical effort for the different gauges discussed in
Sect.\ref{SEC::gauge}, these results strongly suggest that the
spectral gauge is the most favorable choice.

\begin{figure}[t!]
\centering\includegraphics[width=0.45\textwidth]{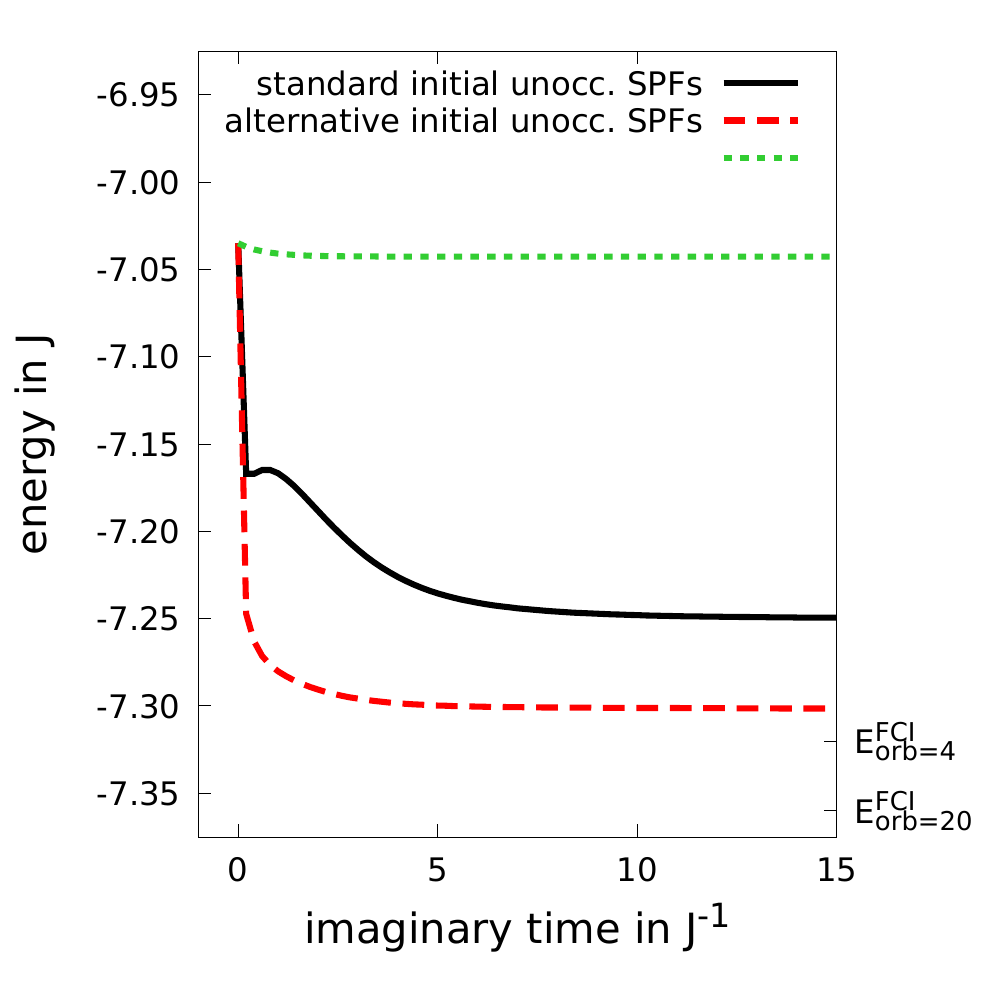}
\caption{\label{Abb::Path} The dependence of the MCTDH-oSQR results
  with $n_{SPF}=2$ and four active orbitals (see
  Fig. \ref{Abb::Gauge-Tree} for the diagrammatic representation of
  the wavefunction) on the choice of the initially unoccupied SPFs is
  shown. The energy is displayed as a function of imaginary time for
  initial unoccupied single particle functions calculated by the
  scheme of Ref.\onlinecite{M8,M10} (solid black line) and for the two
  alternative choices (dash red and dotted green lines) described in
  the text. The energies $E^{FCI}_{orb=20}$ and $E^{FCI}_{orb=4}$ are
  indicated.}
\end{figure}

As already noted above, we found that the MCTDH-oSQR propagation
in imaginary time sometimes does not converge towards the optimal
result for the given number of SPFs. An inspection of the 
wavefunctions appearing in the calculations reveals an 
important property: all single-particle functions show 
distinct local particle numbers which are preserved during the 
entire imaginary time propagation. 

\begin{figure}[t!]
% %trim l u r o
  \centering\includegraphics[width=0.45\textwidth]{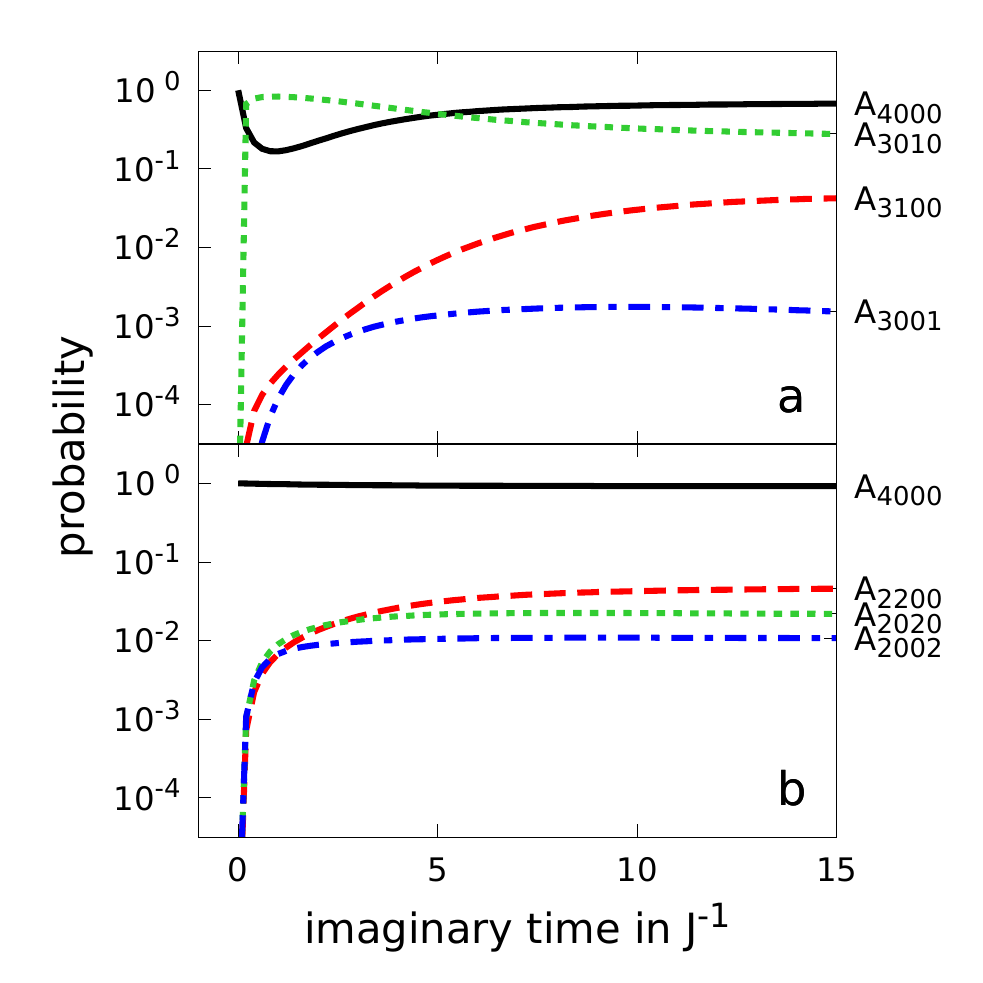}
  \caption{\label{Abb::Acoeff} Non vanishing coefficients
    $A_{i_1..i_4}$ in the MCTDH-oSQR imaginary time
    propagation with $n_{SPF}=2$ and four active orbitals
    (see Fig. \ref{Abb::Gauge-Tree} for the diagrammatic representation 
    of the wavefunction). The choice of the initially unoccupied SPFs
    differs in panel (a) and (b): in panel (a) the 
    scheme of Ref.\onlinecite{M8,M10} is employed while in panel (b) the 
    alternative choice corresponding to the red dashed line in 
    Fig.\ref{Abb::Path} is used. See text for further details.}
\end{figure}
In a single-layer MCTDH-oSQR, the local particle number of the 
SPF $\phi^{(\kappa)}_i$ is defined via the occupation operator
\begin{equation}
\label{ndef1}
\hat{n}^{\kappa}=\hat{b}^\dagger_\kappa \hat{b}_\kappa ~.
\end{equation}
If two degrees of freedom $\nu$ and $\mu$ are combined into a group
$\lambda$ in the ML-MCTDH-oSQR approach, the local particle number for
the group can analogously be defined via the occupation operator
\begin{equation}
\label{ndef2}
\hat{n}^{\lambda} =  \hat{n}^{\nu} + \hat{n}^{\mu} ~.
\end{equation}
The generalization of this concept to an arbitrary tree structure is
straightforward (note that in the following the standard notation
introduced in Ref. \onlinecite{M6} will be used to specify objects
appearing in the multi-layer approach). Thus, the statement that a SPF
$\phi^{\lambda}_i$ shows a distinct local particle number
$n^{\kappa}_i$ implies
\begin{equation}
\hat{n}^{\lambda} \phi^{\lambda}_i = n^{\lambda}_i \phi^{\lambda}_i ~.
\end{equation}

The local particle number conservation can be related to the
convergence problems mentioned above. To this end, the ground state of
the Bose-Hubbard model was calculated employing four orbitals and two
SPFs per orbital (corresponding to $N=2$ in
Fig. \ref{Abb::Gauge-Tree}).  The spectral gauge and the same initial
orbitals as in the above calculations have been employed. Orbital 1 is
initially populated by four bosons (${n}^{1}=4$) while orbitals 2 to 4
are unoccupied (${n}^{2}={n}^{3}={n}^{4}=0$).  The procedure of
Refs.~\onlinecite{M8,M10} used above then yields an unoccupied SPF
with the local particle number ${n}^{1}=3$ associated with the orbital
1 and unoccupied SPFs with local particle numbers ${n}^{\lambda}=1$
associated with the unoccupied orbitals $\lambda=2,3,4$. The result
for this choice of initial SPFs has been depicted
Fig. \ref{Abb::Gauge}.  For comparison, two alternative sets of
choices for the initial unoccupied SPFs have be considered. Here the
local particle number of the unoccupied SPF associated with the
orbital 1 is ${n}^{1}=k$ with $k=1$ or $k=2$. The corresponding
choices for local particle number of the unoccupied SPFs associated
with the unoccupied orbitals $\lambda=2,3,4$ are ${n}^{\lambda}=4-k$.

\begin{figure*}[t!]
\centering\includegraphics[width=0.9\textwidth]{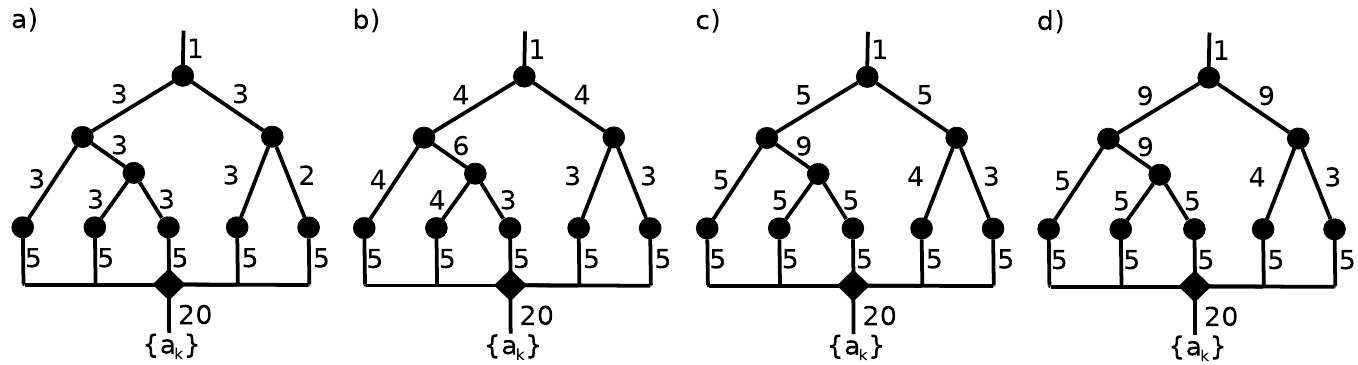}
\caption{\label{Abb::Layered-Tree} Diagrammatic representations of
  the different wavefunctions employed in the ML-\ac{MCTDH-oSQR} calculations
  are shown in panels (a) to (d).}
\end{figure*}
The result of imaginary time propagations starting from the three different
initial choices of the unoccupied SPFs are shown in
Fig. (\ref{Abb::Path}). In the figure, the energy is displayed as a 
function of the imaginary time and the $E^{FCI}_{orb=4}$ and $E^{FCI}_{orb=20}$
values obtained from previous calculations are indicated.
Most strikingly, a strong dependence of the computed energy 
on the initial choice of the unoccupied SPFs is observed. 
The alternative choice $k=1$ (dotted green line) yields an energy
curve which shows almost no decay while the alternative choice $k=2$ 
yields strongly decaying energies converging to a final value close 
to $E^{FCI}_{orb=4}$, the $n_{SPF} \to \infty$ limit for four active orbitals
found in Fig. \ref{Abb::Gauge}.
Remarkably, results obtained with the alternative choice $k=2$ for the
unoccupied SPFs converges to a lower energy than the result obtained
with the ``optimal'' choice for the unoccupied SPFs according to
Refs.~\onlinecite{M8,M10}.

Further insights can be gained by a detailed analysis of the
MCTDH-oSQR wavefunctions obtained by the standard and the alternative
($k=2$) choice of the unoccupied initial SPFs. In both cases the
MCTDH-oSQR wavefunction (as defined in
Eqs.~(\ref{EQ::ImagWF})-(\ref{EQ::ImagWF3}) shows only four
non-vanishing A-coefficients in the CI expansion
(Eq.~(\ref{EQ::ImagWF2})): the wavefunction takes the form
\begin{eqnarray}    
\left| \Psi(\tau)\right> &=& A_{4000}(\tau) \left| \Phi_{4000}(\tau) \right>
+ A_{3100}(\tau) \left| \Phi_{3100}(\tau) \right> \nonumber\\&&
+ A_{3010}(\tau) \left| \Phi_{3010}(\tau) \right>
+ A_{3001}(\tau) \left| \Phi_{3001}(\tau) \right> \nonumber\\&&
\label{A_standard}
\end{eqnarray}    
for the standard choice of the unoccupied initial SPFs and 
\begin{eqnarray}    
\left| \Psi (\tau)\right> &=& A_{4000}(\tau) \left| \Phi_{4000}(\tau) \right>
+ A_{2200}(\tau) \left| \Phi_{2200}(\tau) \right> \nonumber\\&&
+ A_{2020}(\tau) \left| \Phi_{2020}(\tau) \right>
+ A_{2002}(\tau) \left| \Phi_{2002}(\tau) \right> \nonumber\\&&
\label{A_alternative}
\end{eqnarray}    
for the alternative choice of the unoccupied initial SPFs.  The
observed structures in the wavefunctions are a consequence of the
local particle number conservation and the respective choice of the
initial SPFs. The squared absolute values of the non-vanishing
$A_{ijkl}$, i.e., the probabilities to occupied a configuration with
the local particle numbers $i,j,k,l$ in the four orbitals, are shown
in Fig. \ref{Abb::Acoeff} as a function of the imaginary time $\tau$.

The results shown in Fig.~\ref{Abb::Acoeff} provide the insights which allow one
to understand to data displayed in Figs. (\ref{Abb::Path}).
The ``optimal'' choice for the unoccupied SPFs introduced in
Ref.\onlinecite{M8} is actually based a time-local definition
of ``optimal'' referring to the initial time $\tau=0$. Populating a
second configuration with the local particle numbers $3,0,1,0$ 
yields a strong initial decrease of the energy in accordance with
the time-local measure of the ``optimal'' choice of unoccupied SPFs
used in Refs.\onlinecite{M8}. While this choice provides
favorable results for short propagation times $\tau$, it does not
allow for convergence to the optimal result for $\tau \to \infty$.
In contrast, the wavefunction structure resulting from the alternative 
choice of the initial SPFs, Eq.(\ref{A_alternative}), facilitates 
convergence towards lower final energies once the orbitals have been
adapted. While the interplay between SPF and orbital optimization
allows for a smooth exponential energy decay for the alternative choice,
a more complex relaxation behavior is found for the standard
choice in Figs. \ref{Abb::Path} and \ref{Abb::Acoeff}.

Based on the detailed understanding of the MCTDH-oSQR approach 
obtained above, its multi-layer extension will be studied
in the following. The same Hamiltonian and initial wavefunction as in 
the previous MCTDH-oSQR calculations is used. Five orbitals and 
the wave function representations diagrammatically depicted in 
Fig. \ref{Abb::Layered-Tree} are employed. The unoccupied initial 
SPFs have been individually chosen based on results of test calculations 
to assure optimal results. A two step procedure is employed to efficiently compute
the ground state energy. First, the SPFs are propagated in imaginary time using 
frozen orbitals. Then the SPFs and orbitals are simultaneously propagated in 
imaginary time until convergence.

\begin{figure}[t!]
% %trim l u r o
  \centering\includegraphics[width=0.45\textwidth]{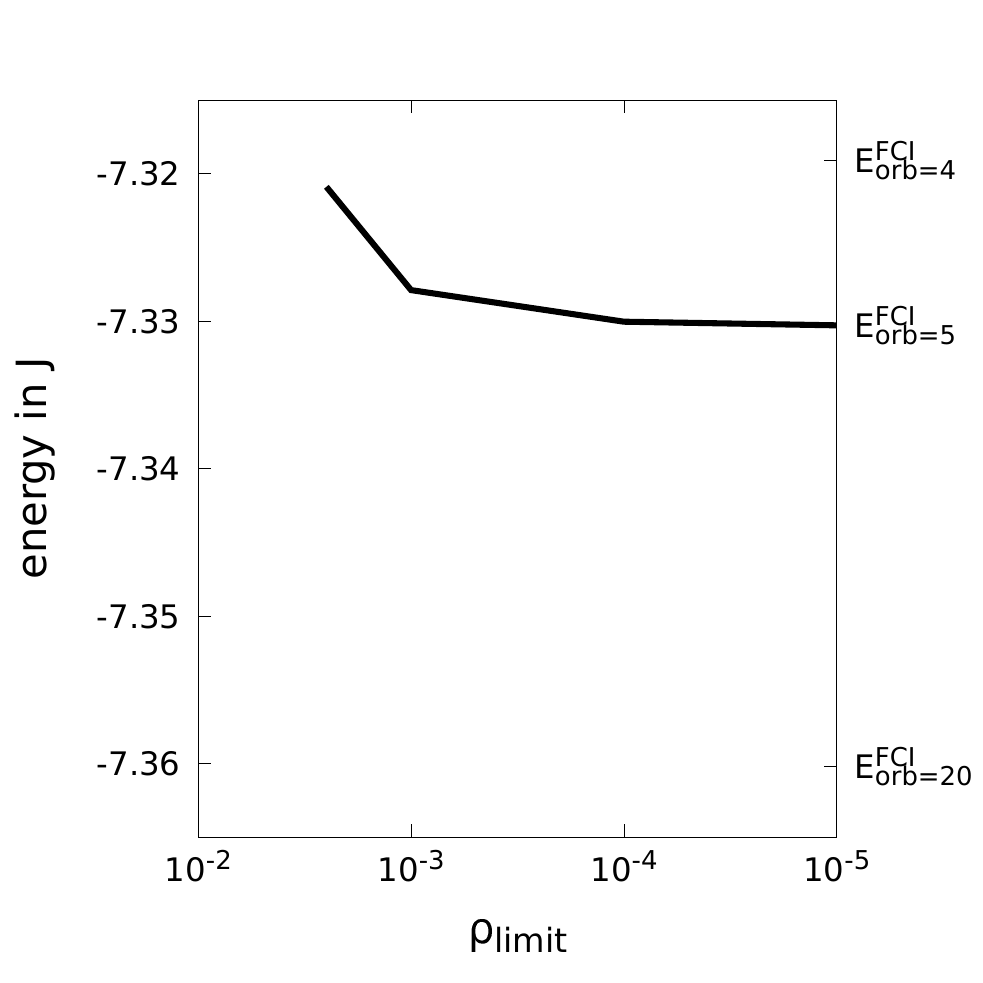}
  \caption{\label{Abb::Orbital_Connvergence} Convergence of the 
   ML-MCTDH-oSQR calculation with respect to the number of SPFs:
   the computed ground state energy is displayed as a function of
   $\rho_{limit}$ (see text for details). 
   The diagrammatic representation of the different
   wavefunctions is given by Fig. (\ref{Abb::Layered-Tree}). 
   The energies $E^{FCI}_{orb=20}$ and $E^{FCI}_{orb=4}$ are indicated.}
\end{figure}

To investigate the convergence of the ML-MCTDH-oSQR approach with
respect to the number of SPFs for a fixed number of orbitals,
different SPF basis set sizes have been considered according to
different upper limits $\rho_{limit}$ for the population of the 
least populated natural SPF present in the wavefunction representation. 
The SPF bases given in panels (a), (b), (c), (d) of 
Fig. \ref{Abb::Layered-Tree} guarantee that the smallest natural population 
is below $2.5\cdot10^{-3}$, $1\cdot10^{-3}$, $1\cdot10^{-4}$, $1\cdot10^{-5}$,
respectively. The ground state energies obtained in the calculations
are plotted in Fig.\ref{Abb::Orbital_Connvergence} as a function of the
$\rho_{limit}$ value employed.  The energy $E^{FCI}_{orb=5}$ computed
with a extremely well converged SPF basis and five active orbitals as
well as the previously calculated energies $E^{FCI}_{orb=4}$ and 
$E^{FCI}_{orb=20}$ are indicated in the figure.

Roughly exponential convergence of the calculated energies towards 
$E^{FCI}_{orb=5}$ as a function of $\rho_{limit}$ is found in 
Fig.\ref{Abb::Orbital_Connvergence}. The smallest basis considered,
which corresponds to $\rho_{limit}=2.5\cdot10^{-3}$, already yields an 
energy below  $E^{FCI}_{orb=4}$, the best result obtainable with 4 active
orbitals and almost converged results can be obtained with the basis 
corresponding to $\rho_{limit}=1\cdot10^{-3}$. It is interesting to
compare the effects resulting from the limited size of SPF bases and
the limited number of active orbitals considered. As indicated in 
Fig.\ref{Abb::NatOrbSQR}, the population of the fifth natural orbital is 
$1.1\cdot10^{-2}$. The present results show that restricting
the number of active orbitals to four and neglecting the fifth
orbital results in a slightly larger error than using SPF basis sets 
corresponding to $\rho_{limit}=2.5\cdot10^{-3}$. Thus, the respective
populations of the natural orbitals and SPFs can be used to roughly
estimate the relative importance of errors resulting from the limited 
numbers of orbitals and SPFs considered.

\section{Discussion}

A central result of the preceding section is the observation of a
local particle number conservation in the individual SPFs employed in
the single-layer and multi-layer MCTDH wavefunctions. This local
particle number conservation has severe consequences. On one hand, it
constrains the space available for the wavefunction during the
imaginary time propagation depending on the initial choice of the
individual SPFs.  This feature severely complicates the convergence
behavior. On the other hand, the local particle number conservation
results in large numbers of vanishing $A$-coefficients and could be
employed to reduce the numerical effort. In this section, we first
provide an analytically derivation of the local particle number
conservation observed in the numerical study. Second, a revised
ML-MCTDH-oSQR scheme which explicitly considers local particle number
conservation in the multi-layer wavefunction representation will be
suggested.

\label{SEC::Discussion}
\subsection{Local Particle Number Conservation}

The local particle number conservation of the single particle functions 
in (multi-layer) MCTDH-SQR type approaches observed in the numerical 
study can be derived analytically. To this end, 
a particle number conserving Hamiltonian
\begin{align}
  [\hat H,\hat N]&=0,
\end{align}
where $\hat N$ denotes the particle number operator,
\begin{align}
  \hat N &= \sum_{k=1}^F \hat a_k^\dagger \hat a_k = \sum_{\mu=1}^F \hat a_\mu^\dagger \hat a_\mu ~,
\end{align}
and a wavefunction $\left|\Psi\right>$ containing $N$ particle,
\begin{align}
  \hat N \Psi &=  N \Psi ~,
\end{align}
is considered. The local particle number operator $\hat{n}^{\lambda}$ 
associated with an arbitrary SPF $\phi^{\lambda}_i$ in the 
multi-layer MCTDH representation has been defined via Eqs. \ref{ndef1} 
and \ref{ndef2}). Single-hole functions $\Psi^\lambda_i$ corresponding
to the SPFs $\phi^{\lambda}_i$ can be defined via
\begin{align}
  \Psi = \sum_i \Psi^\lambda_i \phi^\lambda_i 
\end{align}
and the complementary operator $\hat N^\lambda$ associated with the 
single-hole function $\Psi^\lambda_i$ is given by
\begin{equation}
\hat N^\lambda = \hat N - \hat{n}^{\lambda} ~.
\end{equation}

Using the above definitions, one can obtain equations describing the
action of $\hat N^\lambda$ on the single hole functions and $\hat
n^\lambda$ on the SPFs:
\begin{align}
  \hat N^\lambda \left|\Psi_j^\lambda\right> 
  &= \hat N^\lambda \left<\phi^\lambda_j\middle|\Psi\right>
  \nonumber\\ 
  &= \left<\phi^\lambda_j\middle|\hat N^\lambda \middle|\Psi\right> 
  \nonumber\\ 
  &= \left<\phi^\lambda_j\middle|\hat N- \hat n^\lambda \middle|\Psi\right> 
  \nonumber\\ 
  &= \sum_m \left<\phi^\lambda_j\middle|N- \hat n^\lambda \middle|\phi_m^\lambda\middle>
    \middle|\Psi_m^\lambda\right> 
  \nonumber\\ 
  &= N\left|\Psi_m^\lambda\right> 
  - \sum_m\left<\phi^\lambda_j\middle|\hat n^\lambda\middle|\phi_m^\lambda\middle>
    \middle|\Psi_m^\lambda\right> 
\end{align}
and
\begin{align}
\hat n^\lambda \phi^\lambda_j &=
\hat n^\lambda \left<\Psi^\lambda_j \middle| \Psi \right>
\nonumber\\&=
\left<\Psi^\lambda_j \middle| \hat n^\lambda \middle| \Psi \right>
\nonumber\\&=
\left<\Psi^\lambda_j \middle| \left(\hat N - \hat N^\lambda \right) \middle| \Psi \right>
\nonumber\\&=
\left<\Psi^\lambda_j \middle| \left(N - \hat N^\lambda \right) \middle| \Psi \right>
\nonumber\\&=
N \phi^\lambda_j - \sum_m  \phi^\lambda_m \left<\Psi^\lambda_j \middle|
\hat N^\lambda \middle| \Psi^\lambda_m \right>~.
\end{align}
Thus, $\hat n^\lambda \phi^\lambda_j$ is located in the space spanned 
by the SPFs $\phi^\lambda_m$ and consequently
\begin{align}
  0 = \left(1-\hat P^\lambda\right) \hat n^\lambda \hat \phi^\lambda_j
\label{Eqzero}
\end{align}
where
\begin{equation}
\hat P^\lambda = \sum_m \left| \phi^\lambda_m \right> \left<\phi^\lambda_m\right|
\end{equation}
projects on the space spanned by the SPFs $\phi^\lambda_m$.
 
In the next step the time-derivative of the matrix element
\begin{align}
n^\lambda_{jm} = \left<\phi^\lambda_j\middle|\hat n^\lambda\middle|\phi^\lambda_m \right>
\end{align}
is considered. In the MCTDH approach in second quantization 
representation the local particle number operator take the role of 
coordinate operators and can be considered as time-independent. Thus,
\begin{equation}
\frac{d n^\lambda_{jm}}{dt}
= \left< \frac{\partial \phi^\lambda_j}{\partial t} \middle
|\hat n^\lambda\middle| \phi^\lambda_m \right>
+ \left< \phi^\lambda_j\middle
|\hat n^\lambda\middle|\frac{\partial \phi^\lambda_m}{\partial t} \right> ~.
\label{dndt_1}
\end{equation}
The equation of motion for the SPFs $\phi^\lambda_l$ 
reads \cite{MMC1,M6}
\begin{align}
i \frac{\partial}{\partial t} \phi^{\lambda}_l &=
\sum_m h^{\lambda}_{m l} \phi^\lambda_m \nonumber\\&~~~
+ (1-\hat P^\lambda) \sum_{m,k} (\rho^{\lambda})^{-1}_{lm} 
\left<\Psi^\lambda_m\middle|\hat H\middle|\Psi^\lambda_k \right> \phi^{\lambda}_k
\end{align}
where $h^\lambda$ denotes the gauge matrix and $\rho^\lambda$ the
single-particle density matrix at the respective layer. 
Inserting the above equation into Eq.(\ref{dndt_1}) and employing
Eq.(\ref{Eqzero}), one finds
\begin{align}
\frac{d n^\lambda_{jm}}{dt} & =
i\,\sum_{k,l} \left<\phi^\lambda_l\middle| 
\left<\Psi^\lambda_l\middle|\hat H\middle|\Psi^\lambda_k \right>
(\rho^{\lambda})^{-1}_{kj} (1-\hat P^\lambda) \hat n^\lambda 
\middle|\phi^{\lambda}_m \right>
\nonumber\\
&~-i\,\sum_{k,l} \left<\phi^{\lambda}_j\middle| 
\hat n^{\lambda} (1-\hat P^{\lambda})(\rho^{{\lambda}})^{-1}_{mk} 
\left<\Psi^{\lambda}_k\middle|\hat H\middle|\Psi^\lambda_l \right> 
\middle| \phi^\lambda_l \right>
\nonumber\\
&~+i\,\sum_k(h^{\lambda}_{k j})^* 
\left<\phi^{\lambda}_k\middle|\hat n^{\lambda}\middle| \phi^{\lambda}_m \right>
-i\,\sum_k  h^{\lambda}_{k m} 
\left<\phi^{\lambda}_j\middle|\hat n^{\lambda}\middle| \phi^{\lambda}_k \right>
\nonumber\\
&=i\,\sum_k(h^{\lambda}_{k n})^* n^\lambda_{km}
-i\,\sum_kh^{\lambda}_{k m} n^\lambda_{jk} 
\label{Eq::dercon}
\end{align}
For the standard gauge $h^{\lambda}=0$, which is also employed in the
present work, one finally finds
\begin{align}
&\frac{d}{dt}\left<\phi^{\lambda}_j\middle|\hat n^{\lambda}\middle|\phi^{\lambda}_m \right>=0 ~.
\end{align}
Thus, the representation of the particle number operator in the basis of
SPFs does not change in time. If the initial SPFs are eigenfunctions
of $\hat n^{\lambda}$, the SPFs remain eigenfunctions of 
$\hat n^{\lambda}$ at subsequent points in time.

\subsection{Efficient Tensor Contraction}
\label{SecTenCon}

The conservation of the local particle number can be employed to
devise an approach which combines the tensor contraction
used in ML-MCTDH-SQR and ML-MCTDH-oSQR schemes with the particle
number based truncation scheme used in MCTDHX approaches.

\begin{figure}[t!]
\centering\includegraphics[width=0.2\textwidth]{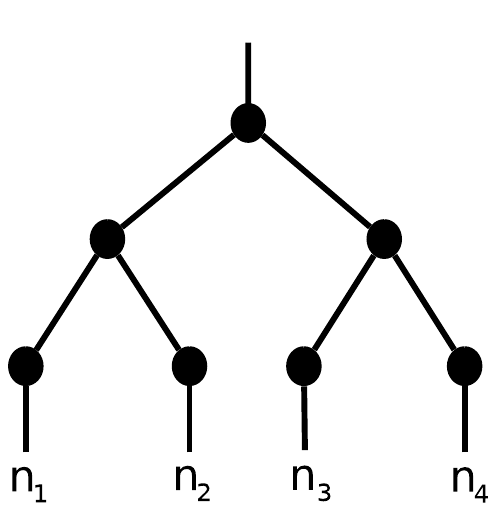}
\caption{\label{Abb::TreeBin} Diagrammatic representation of the
ML-MCTDH-oSQR wavefunction discussed in Sect.\ref{SecTenCon}.}
\end{figure}

The idea of the approach will be illustrated for a simple example: 
a system with constant particle number $N$ described
by the wavefunction representation diagrammatically depicted in
Fig.~\ref{Abb::TreeBin}. Each bottom layer SPF $\phi^\lambda_i$ has 
to show a defined local particle number $n_i$. Thus, the 
bottom layer SPFs read 
\begin{align}
  \phi^{2;1,1}_{i;n_i} &= \frac{(\hat b_1^\dagger)^{n_i}}{\sqrt{n_i}}\left|0\right>, \\
  \phi^{2;1,2}_{i;n_i} &= \frac{(\hat b_2^\dagger)^{n_i}}{\sqrt{n_i}}\left|0\right>, \\
  \phi^{2;2,1}_{i;n_i} &= \frac{(\hat b_3^\dagger)^{n_i}}{\sqrt{n_i}}\left|0\right>, \\
  \phi^{2;2,2}_{i;n_i} &= \frac{(\hat b_4^\dagger)^{n_i}}{\sqrt{n_i}}\left|0\right>.
\end{align}
Note that in addition on the notation introduced in Ref.\onlinecite{M6} 
here the local particle number $n_i$ has be added as an additional subscript. 
The first layer SPFs are then given by
\begin{align}
\phi^{1;1}_{k;n} &= \sum_{\substack{i,j\\n_i+n_j=n}} 
A^{(2;1)}_{k;i,j} \phi^{2;1,1}_{i;n_i} \phi^{2;1,2}_{j;n_j},\\
\phi^{1;2}_{k;n} &= \sum_{\substack{i,j\\n_i+n_j=n}} 
A^{(2;2)}_{k;i,j} \phi^{2;2,1}_{i;n_i} \phi^{2;2,2}_{j;n_j}. 
\end{align}
Here the number of coefficients $A^{(2;1)}_{i,j}$ and $A^{(2;2)}_{i,j}$
which have to be considered is reduced by the local particle number constraint.
On the top layer, the resulting $N$ particle wavefunction can be written as
\begin{align}
  \Psi = \sum_{\substack{i,j\\n_i+n_j=N}} A^1_{i,j} \phi^{1;1}_{i;n_i} \phi^{1;2}_{j;n_j}.
\end{align}
Again, the number of terms in the expansion is reduced by the particle number constraint.

On each layer, single hole functions $\Psi^\lambda_{i;N_i}$ with a defined local 
particle number $N_i$ can be defined via
\begin{align}
  \Psi = \sum_i \Psi^\lambda_{i;N-n_i} \phi^\lambda_{i;n_i} ~. 
\end{align}
The equations of motion for the SPFs on an arbitrary layer which originally read
\begin{equation}
i \dot\phi^\lambda_{i;n_i} 
= (1-\hat P^\lambda) \sum_{j,k} (\rho^\lambda)^{-1}_{ij}
\left<\Psi^\lambda_{j;N_j}\middle|\hat H\middle|\Psi^\lambda_{k;N_k}\right> \phi^\lambda_{k;n_k}
\end{equation}
can then be rewritten as
\begin{equation}
i \dot\phi^\lambda_{i;n_i} 
= (1-\hat P^\lambda) \sum_{\substack{j,k\\n_j=n_i}} (\rho^\lambda)^{-1}_{ij}
\left<\Psi^\lambda_{j;N_j}\middle|\hat H\middle|\Psi^\lambda_{k;N_k}\right> \phi^\lambda_{k;n_k}
\end{equation}
due to particle number conservation. For Hamiltonians which can be written in a 
sum of product form the equations of motion can be further simplified. Here the
Hamiltonian can be cast into the form   
\begin{equation}
\hat{H} = \sum_k c_k \cdot \hat{\mathscr{H}}^{\lambda}_{k;-a_k} \cdot
\hat{h}^{\lambda}_{k;a_k}
\end{equation}
where the operator $\hat{h}^{\lambda}_{k;a}$ acts only on the orbitals contained
in the SPFs $\phi^{\lambda}_{m;n_m}$ and increases the local particle number by $a_k$
while $\hat{\mathscr{H}}^{\lambda}_{k;-a}$ acts only on the orbitals contained
in the single-hole functions $\Psi^{\lambda}_{m;N_m}$ and decreases the particle
number by $a_k$. The equations of motion then read 
\begin{align}
\label{EQ::EOMSPF}
i \dot\phi^\lambda_{i;n_i} 
& = (1-\hat P^\lambda) \sum_k c_k \sum_{\substack{j,m\\n_j=n_i\\n_m=n_i-a_k}} 
(\rho^\lambda)^{-1}_{ij} \nonumber\\ & \cdot
\left<\Psi^\lambda_{j;N_j}\middle|\hat{\mathscr{H}}^{\lambda}_{k;-a_k}|\Psi^\lambda_{m;N_m}\right>\hat{h}^{\lambda}_{k;a_k} \phi^\lambda_{m;n_m} ~.
\end{align}
Thus, for each summand in the Hamiltonian only small subblocks of the
single-hole matrices has to be considered and only specific subsets of the
set of SPFs are coupled in the equations of motion. This significantly reduces
the numerical effort required to solve the equations of motion.

\section{Concluding Remarks}
\label{SEC::Conclusion}

Imaginary time propagation with the recently introduced
\ac{ML-MCTDH-oSQR} approach was studied in the present work. The
\ac{ML-MCTDH-oSQR} approach combines time-dependent optimized orbitals
known from MCTDHX-type approaches with general tensor contractions
known from the ML-MCTDH-SQR approach.  The equations of motions for
imaginary time propagation are derived and differences between the
real and imaginary time propagation are discussed.  An important
difference concerns the gauge operators which determine the rotation
within the active orbital space. The gauge operators used in imaginary
time propagations are anti-hermitian while the gauge-operators used in
real-time propagations are hermitian. This has important consequences
regarding the choice of an efficient 'gauge'. The natural gauge, which
has found to be efficient and yield good result in real time
propagations, scales unfavorably in imaginary time calculations and
should be replaced by the newly introduced spectral gauge. The
spectral gauge is designed to minimize the spectral width of the
effective Hamiltonian.  It shows a numerically favorable scaling in
imaginary time calculations and yields computational results which for
a given basis set size are at least as accurate as the ones obtained
within the natural gauge.

Furthermore, the effects of particle number conservation have been
studied.  The MCTDH-SQR and MCTDH-oSQR approaches are Fock space
methods which do not constrain the particle number by
design. Considering a Bose-Hubbard model as an example of a particle
number conserving system, the implications of particle number
conservation on imaginary time propagation have been studied in
detail. It is found that particle number conservation must be
explicitly ensured to guarantee numerically stable results: during the
imaginary time propagation small occupations of unphysical states can
grow exponentially and finally result in convergence toward states
with incorrect particle numbers. This issue can be resolved by the
introduction of a penalty operator which explicitly increases the
energy of states with an incorrect particle number.
%evt. was gezeigt wurde

Beside the conservation of the total particle number, an interesting
local particle conservation in the tensor networks used in the 
MCTDH-oSQR and ML-MCTDH-(o)SQR approaches has be identified. 
In a second quantization representation, the bottom layer SPFs are 
represented in a particle number basis. If accordingly initialized, 
local particle number can be associated not only with each SPF
on the bottom layer but also with any upper layer SPF. This local 
particle number is a constant of motion. The conservation law has
significant numerical implications. On one hand, it can result in
a complex dependence of the numerical results on the choice of the
initially unoccupied SPFs in the (multi-layer) MCTDH-oSQR approach.
On the other hand, it can be employed to design a new, more efficient
tensor contraction scheme. 

This new tensor contraction scheme combines the advantages of the
MCTDHX and ML-MCTDH-oSQR approaches. It simultaneously uses a 
particle number based truncation scheme and tensor contraction to
reduce the effective size of the tensor network. Furthermore,
the equation of motions for the various elements in the tensor
network show a favorable block structure resulting in a 
potentially very efficient quantum dynamics approach.

\label{SEC::conclusion}

\section*{ACKNOWLEDGMENTS}

The authors thank Sven Krönke and Peter Schmelcher for inspiring
discussions.  The authors are grateful for financial support via the
FCI (Fonds der Chemischen Industrie).

\section*{Appendix A}

Here the equations of motions in imaginary time, given in
Eqs. (\ref{EQ::AEOMIM}),(\ref{EQ::CEOMIM}), are derived. To this end, the
Dirac-Frenkel variational principle is employed. The variation of the
wave function reads
\begin{align}
  \label{Eq:variation}
  \delta \left| \Psi \right>
  &= \sum_{i_1 .. i_f} \left(\delta A_{i_1 .. i_f}(t)\right)\,\left| \Phi_{i_1.. i_f}(\tau)\right>
  + \sum_{\mu=1}^f \left(\delta \hat b_\mu^\dagger\right) \hat b_\mu \left| \Psi \right>.
\end{align}
The time-dependence of the wave function is given by
\begin{align}
  \label{Eq:ddt}
  \frac{d}{d\tau} \left| \Psi \right>
  &= \sum_{i_1 .. i_f} \,\frac{dA_{i_1 .. i_f}}{d\tau}\,\left| \Phi_{i_1.. i_f}(\tau)\right>
  + \sum_{\mu=1}^f\frac{d\hat b_\mu^\dagger}{d\tau}\,\hat b_\mu \left| \Psi \right>.
\end{align}
Inserting Eqs. (\ref{Eq:variation}) and (\ref{Eq:ddt}), into a Dirac
Frenkel variational participle in imaginary time,
\begin{align}
  0=\left< \delta\Psi \middle| \frac{d}{d\tau} + \hat H \middle| \Psi\right>,
\end{align}
yields
\begin{align}
  \label{Eq:VarA}
  0 &= \sum_{i_1 .. i_f} \delta A^*_{i_1 .. i_f} \left(\frac{dA_{i_1 .. i_f}}{d\tau} + \left< \Phi_{i_1 .. i_f} 
      \middle| \hat H + \sum_{\mu=1}^f\frac{d\hat b^\dagger_\mu}{d\tau} \hat b_\mu 
      \middle| \Psi\right> \right) \nonumber\\
  &~~~+ \left<\Psi \middle| \sum_{\mu=1}^f \hat b^\dagger_\mu (\delta \hat b_\mu) 
    \middle(\sum_{i_1 .. i_f}\frac{dA_{i_1 .. i_f}}{d\tau}  \middle| \Phi_{i_1 .. i_f} \right> \nonumber\\
  &~~~~~~~~~~~~~~~~~~~~~~~~~~~~~+ \hat H \left| \Psi\middle> 
    + \sum_{\nu=1}^f\frac{d\hat b^\dagger_\nu}{d\tau} \hat b_\nu \middle| \Psi\middle> \right)
\end{align}
Considering first the variation with respect to $A^*_{i_1 .. i_f}$, one finds
\begin{align}
  0 = \frac{dA_{i_1 .. i_f}}{d\tau} + \left< \Phi_{i_1 .. i_f} \middle| 
    \hat H - \sum_{\mu=1}^f\frac{d\hat b^\dagger_\mu}{d\tau} \hat b_\mu \middle| \Psi\right>.
\end{align}
In this equation $\frac{d\hat b_\mu^\dagger}{d\tau}$ can be replaced by
$-\sum_{l=1}^F \hat b^\dagger_l \bar h_{l \mu}$ (see
Eq. (\ref{EQ::dbdagger})).  Here $l$ runs from $1$ to $F$, but
$\left|\Psi\right>$ contains only creators $\hat b^\dagger_\nu$ with index
$\nu=1,..,f$. Therefore $l$ can be replaced by an index $\nu$ running
from $1$ to $f$. Thus, the time evolution of $A_{i_1 .. i_f}$ is given by
\begin{align}
  \label{EQ::DiffA}
  -\frac{dA_{i_1 .. i_f}}{d\tau} = \left< \Phi_{i_1 .. i_f} \middle| 
    \hat H - \sum_{\nu, \mu = 1}^f \hat b^\dagger_\nu \bar h_{\nu \mu} \hat b_\mu \middle| \Psi\right>.
\end{align}

The equations of motions for the orbitals $c_{\mu k}$ are derived from Eq.(\ref{Eq:VarA}) 
considering the variation with respect to $\hat b_{\mu}$:
\begin{align}
  0=&\left<\Psi \middle| \sum_{\mu=1}^f \hat b^\dagger_\mu (\delta \hat b_\mu) 
    \middle(\sum_{i_1 .. i_f}\frac{dA_{i_1 .. i_f}}{d\tau}  \middle| \Phi_{i_1 .. i_f} \right> 
\nonumber\\ & ~~~~~~~~~~~~~~~~~~~~~~~~~
+ \hat H \left| \Psi\middle> 
  + \sum_{\nu=1}^f\frac{d\hat b^\dagger_\nu}{d\tau} \hat b_\nu \middle| \Psi\middle> \right)
\end{align}
Using
\begin{align}
  \delta \hat b_\mu &= \delta c^*_{\mu k} \hat a_k
\end{align}
and Eq.(\ref{EQ::dbdagger}) one finds
\begin{align}
  0=&\sum_{\mu=1}^f \sum_{k=1}^F(\delta c^*_{\mu k}) \left<\Psi \middle| 
    \hat b^\dagger_\mu \hat a_k \middle(\sum_{i_1 .. i_f}\frac{dA_{i_1 .. i_f}}{d\tau}  
    \middle| \Phi_{i_1 .. i_f} \right> 
  \nonumber\\ 
  & ~~~~~~~~~~~~~~~~~~~~~~~~+ \hat H \left| \Psi\middle> 
    + \sum_{\nu = 1}^f \sum_{l = 1}^F \frac{dc_{\nu l}}{d\tau} \hat a^\dagger_l \hat b_\nu 
    \middle| \Psi\middle> \right)
\end{align}
Thus, the equations
\begin{align}
  0=&\left<\Psi \middle| \hat b^\dagger_\mu \hat a_k \middle(\sum_{i_1 .. i_f}
    \frac{dA_{i_1 .. i_f}}{d\tau}  \middle| \Phi_{i_1 .. i_f} \right> 
  \nonumber\\ & ~~~~~~~~~~~~~~~~~~~~~~~~~
  + \hat H \left| \Psi\middle> + \sum_{\nu = 1}^f \sum_{l = 1}^F 
    \frac{dc_{\nu l}}{d\tau} \hat a^\dagger_l \hat b_\nu \middle| \Psi\middle> \right)
\end{align}
have to be satisfied for all $\mu=1,..,f$ and $k=1,..,F$.

Inserting Eq. (\ref{EQ::DiffA}), the above equations can be rewritten as
\begin{align}
\label{EQ:SUPORT1}
0=
&\left<\Psi \middle| \hat b^\dagger_\mu \hat a_k \middle(1 
- \sum_{i_1 .. i_f}  \middle| \Phi_{i_1 .. i_f} \middle>\middle< \Phi_{i_1 .. i_f} \middle| \right) 
\nonumber\\ &~~~~~~~~~~~~
\cdot\left(\hat H + \sum_{\nu = 1}^f \sum_{l = 1}^F \frac{dc_{\nu l}}{d\tau} 
  \hat a^\dagger_l \hat b_\nu \middle)\middle| \Psi\right> 
\end{align}
At this point it is important to note that 
\begin{align}
\hat{P} = \sum_{i_1 .. i_f}  \left| \Phi_{i_1 .. i_f} \middle>
\middle< \Phi_{i_1 .. i_f} \right|
\end{align}
is the projector onto a space spanned by all possible occupations of active orbitals 
included in the occupations number space $\{i_1,..,i_f\}$. If one assumes that this
space is chosen sufficiently large to ensure that
\begin{align}
\hat{P} \hat{b}^\dagger_\nu \left|\Psi\right> &=  
\hat{b}^\dagger_\nu \left|\Psi\right> 
\end{align}
for all $\nu=1,..,f$, the equations can be further simplified. Then one finds
\begin{align}
\left(1-\hat{P}\right) \hat{a}^\dagger_k \hat{b}_\mu \left|\Psi\right> &=
\left(1-\hat{P}\right) \sum_{l=1}^F c_{l k}^* \hat{b}^\dagger_l \hat{b}_\mu \left|\Psi\right>
\nonumber\\&=
\sum_{l=f+1}^F c_{l k}^* \hat{b}^\dagger_l \hat{b}_\mu \left|\Psi\right>
\nonumber\\&=
\left(\hat{a}^\dagger_k
- \sum_{\lambda=1}^f c_{\lambda k}^* \hat{b}^\dagger_\lambda \right)
\hat{b}_\mu \left|\Psi\right>
\nonumber\\&=
\left(\hat{a}^\dagger_k
- \sum_{\lambda=1}^f \sum_{l=1}^F c_{\lambda k}^* c_{\lambda l} \hat{a}^\dagger_l \right)
\hat{b}_\mu \left|\Psi\right>
\nonumber\\&=
\sum_{l=1}^F \left(\delta_{kl}
- \sum_{\lambda=1}^f \sum_{l=1}^F c_{\lambda k}^* c_{\lambda l} \right)
\hat{a}^\dagger_l \hat{b}_\mu \left|\Psi\right>
\end{align}
Inserting the above equation into Eq. (\ref{EQ:SUPORT1}) yields
\begin{align}
0&= 
\sum_{l=1}^F \left(\delta_{k l} - \sum_{\lambda=1}^f c_{\lambda k} c^*_{\lambda l} \right)
\nonumber\\&~~~~~~~~~
\cdot\left<\Psi \middle| \hat b^\dagger_\mu \hat a_l
\middle(\hat H  + \sum_{\nu = 1}^f \sum_{n = 1}^F \frac{dc_{\nu n}}{d\tau} 
\hat a^\dagger_n \hat b_\nu \middle)\middle| \Psi\right>.
\end{align}
The above equation can be further simplified by commuting $\hat a_l$ to the left:
\begin{align}
0&=
\sum_{l=1}^F \left(\delta_{k l} - \sum_{\nu=1}^f c_{\nu k} c^*_{\nu l} \right)
\nonumber\\ &~~~
\cdot\left\{\left<\Psi \middle| \hat b^\dagger_\mu \left[\hat a_l, \hat H
+ \sum_{\lambda = 1}^f \sum_{n=1}^F\frac{dc_{\lambda n}}{d\tau} \hat a_n^\dagger \hat b_\lambda \right]
\middle| \Psi\right>\right.
\nonumber\\ &~~~~~~
+ \left.\left<\Psi \middle| \hat b^\dagger_\mu \left(\hat H
+ \sum_{\lambda = 1}^f \sum_{n=1}^F\frac{dc_{\lambda n}}{d\tau} \hat a_n^\dagger \hat b_\lambda \right) \hat a_l
\middle| \Psi\right>\right\}
\end{align}
Using the relation $\sum_{l=1}^F \left(\delta_{k l} - \sum_{\nu=1}^f
  c_{\nu k} c^*_{\nu l} \right) \hat a_l \left| \Psi\right> = 0$ yields
\begin{align}
0&=
\sum_{l=1}^F \left(\delta_{k l} - \sum_{\nu=1}^f c_{\nu k} c^*_{\nu l} \right)
\nonumber\\ &~~~
\cdot\left<\Psi \middle| \hat b^\dagger_\mu \left[\hat a_l, \hat H
+ \sum_{\lambda = 1}^f \sum_{n=1}^F\frac{dc_{\lambda n}}{d\tau} \hat a_n^\dagger \hat b_\lambda \right]
\middle| \Psi\right>
\nonumber\\&=
\sum_{l=1}^F \left(\delta_{k l} - \sum_{\nu=1}^f c_{\nu k} c^*_{\nu l} \right)
\nonumber\\ &~~~
\cdot\left<\Psi \middle| \hat b^\dagger_\mu [\hat a_l, H]
+ \sum_{\lambda = 1}^f \frac{dc_{\lambda l}}{d\tau} \hat b^\dagger_\mu \hat b_\lambda\middle| \Psi\right>
\end{align}
If the Hamiltonian $\hat H$ is in normal order the all operators $\hat a_i$ can
be replaced by active operators $\hat b_\kappa$:
\begin{align}
0&=
\sum_{l=1}^F \left(\delta_{k l} - \sum_{\nu=1}^f c_{\nu k} c^*_{\nu l} \right)
\nonumber\\ &~~~
\cdot\left<\Psi \middle| \hat b^\dagger_\mu [\hat a_l, \hat H]_{\hat a_i=\sum\limits_{\kappa=1}^f c_{\kappa i} \hat b_\kappa}
+ \sum_{\lambda = 1}^f \frac{dc_{\lambda l}}{d\tau} \hat b^\dagger_\mu \hat b_\lambda\middle| \Psi\right>
\end{align}

Applying the projector to each term, one finds
\begin{align}
0&=\sum_{\mu=1}^f \sum_{k=1}^F (\delta c^*_{\mu k}) 
\nonumber\\ &\cdot
\left\{\sum_{l=1}^F \left(\delta_{k l} - \sum_{\nu=1}^f c_{\nu k} c^*_{\nu l} \right)
\left<\Psi \middle| \hat b^\dagger_\mu \left[\hat a_l,\hat H\right]_{\hat a_i=\sum\limits_{\kappa=1}^f c_{\kappa i} \hat b_\kappa}\middle| \Psi\right> 
\right.\nonumber\\ &~~ \left.
+ \sum_{\nu = 1}^f \frac{dc_{\nu k}}{d\tau} \left<\Psi \middle| \hat b^\dagger_\mu \hat b_\nu \middle| \Psi\right> 
- \sum_{\nu = 1}^f \bar h_{\nu \lambda} \left<\Psi \middle| \hat b^\dagger_\mu \hat b_\lambda \middle| \Psi\right>
\right\}.
\end{align}
The resulting equation of motion for the orbitals $c_{\mu k}$ reads:
\begin{align}
&-\frac{dc_{\mu k}}{d\tau}=
\sum_{\nu = 1}^f \bar h_{\mu \nu} c_{\nu k}
+  \sum_{l=1}^F \left(\delta_{k l} - \sum_{\nu=1}^f c_{\nu k} c^*_{\nu l} \right) 
\nonumber\\ & ~~~~~~~~
\cdot\sum_{\lambda=1}^f \rho^{-1}_{\mu \lambda}
\left<\Psi \middle| \hat b^\dagger_\lambda \left[\hat a_l,\hat H\right]_{\hat a_i=\sum\limits_{\kappa=1}^f c_{\kappa i} \hat b_\kappa}\middle| \Psi\right>
\end{align}

\section*{Appendix B}

The properties of the spectral gauge can be illustrated by a simple
example, the one-particle Hamiltonian
\begin{align}
\hat H = \sum_{k,l = 1}^F w_{k l}\, \hat b^\dagger_k \hat b_l.
\end{align}
For simplicity the analysis is presented in natural orbitals (see
Eqs. (\ref{EQ::NatRepa})-(\ref{EQ::NatRepd})). Then the Hamiltonian
reads
\begin{align}
\hat H = \sum_{k,l = 1}^F \tilde w_{k l}\, \hat{\tilde b}^\dagger_k \hat{\tilde b}_l
\end{align}
where
\begin{align}
\tilde w_{r s} = \sum_{k,l = 1}^F u^*_{k r} w_{k l} u_{s l}
\end{align}
The corresponding spectral gauge matrix is
\begin{align}
  \tilde h^{(spec)}_{\xi \sigma} &= \sum_{k,l = 1}^F \tilde w_{k l}
  \frac{\left<\Psi\middle|\left[\hat{\tilde b}^\dagger_k \hat{\tilde
          b}_l ,\hat{\tilde b}^\dagger_\sigma \hat {\tilde b}_\xi\right]
      \middle|\Psi\right>} {\tilde
    \rho_{\xi}+ \tilde \rho_{\sigma} + \epsilon}\\
  &= \sum_{k,l = 1}^F \tilde w_{k l} \frac{\delta_{l \sigma}
    \left<\hat{\tilde b}_k \Psi\middle| \hat{\tilde b}_\xi \Psi\right> -
    \delta_{k \xi} \left<\hat{\tilde b}_\sigma \Psi\middle| \hat{\tilde
      b}_l \Psi\right>} {\tilde \rho_{\xi}+ \tilde \rho_{\sigma} +
    \epsilon}.
\end{align}
Since $\hat{\tilde b}_k \left|\Psi\right>=0$ for $k=f+1, ..., F$, one finds
\begin{align}
  \tilde h^{(spec)}_{\xi \sigma} &= 
\frac{\sum\limits_{\mu = 1}^f \tilde w_{\mu \sigma} \left<\hat{\tilde b}_\mu \Psi\middle| \hat{\tilde b}_\xi \Psi\right>
-\sum\limits_{\mu = 1}^f \tilde w_{\xi \mu} \left<\hat{\tilde b}_\sigma \Psi\middle| \hat{\tilde b}_\mu \Psi\right>}
 {\tilde \rho_{\xi}+ \tilde \rho_{\sigma} + \epsilon}.
\end{align}
Using Eq. (\ref{EQ::NatCon}), one arrives at
\begin{align}
  \tilde h^{(spec)}_{\xi \sigma} &= \tilde w_{\xi \sigma}
\frac{\tilde \rho_\xi - \tilde \rho_\sigma} {\tilde
    \rho_{\xi}+ \tilde \rho_{\sigma} + \epsilon}.
\end{align}
The eigenstates of an one-particle operator show only one occupied
natural orbital and a diagonal $\tilde w_{kl}$. Thus, $\tilde
h^{(spec)}_{\xi \sigma}$ vanishes in the limit of a converged
calculation. Furthermore, the action of $h^{(spec)}$ will ensure
rotation of the orbital basis towards natural orbitals in course of
the imaginary time propagation.

\bibliographystyle{jcp}
%\bibliography{literatur,eigen,MCTDH,MCTDHF,MCTDHB,methods,sonstig}
%\bibliography{/raid/home/weike/Literatur/literatur.bib}

\end{document}